\documentclass[journal]{IEEEtran}
\usepackage{cite, bm,amsmath,graphicx,amssymb,mathtools,amsthm,subcaption,footmisc,framed,booktabs,dsfont,array,multirow,makecell}
\usepackage{mathrsfs}
\usepackage{epstopdf}
\usepackage{threeparttable}
\usepackage{tabularx}
\usepackage{algorithm}
\usepackage{algpseudocode}
\usepackage{xcolor}
\algnewcommand\Input{\item[\textbf{Input:}]}
\usepackage[english]{babel}
\addto\captionsenglish{\renewcommand{\figurename}{Fig.}}
\usepackage{url}
\usepackage[colorlinks=true,linkcolor=black,citecolor=black,urlcolor=blue]{hyperref}
\begin{document}
\newcommand{\thd}[1]{$#1$-th}
\newcommand{\dt}[1]{d_{\bm{t};{{#1}}}}
\newcommand{\du}[1]{d_{{{#1}};\bm{r}}}
\newcommand{\wn}{\frac{2\pi}{\lambda}}
\newcommand{\expect}{\operatorname{\mathbb{E}}}
\newcommand{\ch}{\mathbf{h}}
\newcommand{\Cq}{ C_\theta }
\newcommand{\Cr}{C_r }
\newcommand{\diag}{\operatorname{diag}}
\newcommand{\deltaT}{\Delta \hat{\theta}}
\newcommand{\epsT}{\varepsilon_\theta}
\newcommand{\epsL}{\varepsilon_r}
\newcommand{\snr}{{\rm SNR}}
\newcommand{\cod}{\Delta C(\hatT,\theta,\hatD,r)}
\newcommand{\Capp}{{{C}}}
\newcommand{\cae}{{\mathcal{E}}}
\newcommand{\bfa}{{\mathbf{a}}}
\newcommand{\codew}{{\boldsymbol{w}}}
\newcommand{\rrg}{r_{\rm rg}}
\newcommand{\lrg}{\ell_{\rm rg}}
\newcommand{\minimize}{\operatorname*{minimize}}
\newcommand{\trg}{\theta_{\rm rg}}
\newcommand{\rrgq}{\rho_{\rm rg}}
\newcommand{\cFullFun}{C(\epsL,\epsT)}
\newcommand{\frf}{\mathbf{F}_{\mathrm{RF}}}
\newcommand{\fbb}{\mathbf{F}_{\mathrm{BB}}}
\newcommand{\fbbk}[1]{\mathbf{f}_{\mathrm{BB},#1}}
\newcommand{\frfk}[1]{\mathbf{f}_{\mathrm{RF},#1}}
\newcommand{\Bcal}{\mathcal{B}}
\newcommand{\Scal}{\mathcal{S}}

\newtheoremstyle{custom_style} {\topsep} {\topsep} {\normalfont} {} {\bfseries} {} {.2em} {\thmname{#1} {\normalfont\textbf{\thmnumber{#2}}} \thmnote{\normalfont(#3)\textbf{.}}} 

\theoremstyle{custom_style}

\newtheorem{theorem}{\emph{\underline{Theorem}}}
\newtheorem{acknowledgement}[theorem]{Acknowledgement}
\newtheorem{axiom}[theorem]{Axiom}
\newtheorem{case}[theorem]{Case}
\newtheorem{claim}[theorem]{Claim}
\newtheorem{conclusion}[theorem]{Conclusion}
\newtheorem{condition}[theorem]{Condition}
\newtheorem{conjecture}[theorem]{\emph{\underline{Conjecture}}}
\newtheorem{criterion}[theorem]{Criterion}
\newtheorem{definition}{\emph{\underline{Definition}}}
\newtheorem{exercise}[theorem]{Exercise}
\newtheorem{lemma}{\emph{\underline{Lemma}}}
\newtheorem{example}{\emph{\underline{Example}}}
\newtheorem{observation}{\emph{\underline{Observation}}}
\newtheorem{corollary}{\emph{\underline{Corollary}}}
\newtheorem{notation}[theorem]{Notation}
\newtheorem{problem}[theorem]{Problem}
\newtheorem{proposition}{\emph{\underline{Proposition}}}
\newtheorem{solution}[theorem]{Solution}
\newtheorem{method}[theorem]{Method}
\newtheorem{summary}[theorem]{Summary}
\newtheorem{assumption}{\emph{\underline{Assumption}}}
\newtheorem{remark}{\bf \emph{\underline{Remark}}}

\def\qed{$\Box$}
\def\QED{\mbox{\phantom{m}}\nolinebreak\hfill$\,\Box$}
\def\endproof{\hspace*{\fill}~\qed\par\endtrivlist\vskip3pt}

\def\E{\mathsf{E}}
\def\eps{\varepsilon}
\def\Lsp{{\boldsymbol L}}
\def\Bsp{{\boldsymbol B}}
\def\lsp{{\boldsymbol\ell}}
\def\Ltsp{{\Lsp^2}}
\def\Lpsp{{\Lsp^p}}
\def\Linsp{{\Lsp^{\infty}}}
\def\LtR{{\Lsp^2(\Rst)}}
\def\ltZ{{\lsp^2(\Zst)}}
\def\ltsp{{\lsp^2}}
\def\ltZt{{\lsp^2(\Zst^{2})}}
\def\ninN{{n{\in}\Nst}}
\def\oh{{\frac{1}{2}}}
\def\grass{{\cal G}}
\def\ord{{\cal O}}
\def\dist{{d_G}}
\def\conj#1{{\overline#1}}
\def\ntoinf{{n \rightarrow \infty }}
\def\toinf{{\rightarrow \infty }}
\def\tozero{{\rightarrow 0 }}
\def\trace{{\operatorname{trace}}}
\def\ord{{\cal O}}
\def\UU{{\cal U}}
\def\rank{{\operatorname{rank}}}
\def\acos{{\operatorname{acos}}}

\def\SINR{\mathsf{SINR}}
\def\SNR{\mathsf{SNR}}
\def\SIR{\mathsf{SIR}}
\def\tSIR{\widetilde{\mathsf{SIR}}}
\def\Ei{\mathsf{Ei}}
\def\l{\left}
\def\r{\right}
\def\({\left(}
\def\){\right)}
\def\llargebra{\left\{}
\def\rlargebra{\right\}}
\def\lmidbra{\left[}
\def\rmidbra{\right]}
\def\labs{\left|}
\def\rabs{\right|}

\setcounter{page}{1}

\newcommand{\eref}[1]{(\ref{#1})}
\newcommand{\fig}[1]{Fig.\ \ref{#1}}

\def\bydef{:=}
\def\ba{{\mathbf{a}}}
\def\bb{{\mathbf{b}}}
\def\bc{{\mathbf{c}}}
\def\bd{{\mathbf{d}}}
\def\bee{{\mathbf{e}}}
\def\bff{{\mathbf{f}}}
\def\bg{{\mathbf{g}}}
\def\bh{{\mathbf{h}}}
\def\bi{{\mathbf{i}}}
\def\bj{{\mathbf{j}}}
\def\bk{{\mathbf{k}}}
\def\bl{{\mathbf{l}}}
\def\bm{{\mathbf{m}}}
\def\bn{{\mathbf{n}}}
\def\bo{{\mathbf{o}}}
\def\bp{{\mathbf{p}}}
\def\bq{{\mathbf{q}}}
\def\br{{\mathbf{r}}}
\def\bs{{\mathbf{s}}}
\def\bt{{\mathbf{t}}}
\def\bu{{\mathbf{u}}}
\def\bv{{\mathbf{v}}}
\def\bw{{\mathbf{w}}}
\def\bx{{\mathbf{x}}}
\def\by{{\mathbf{y}}}
\def\bz{{\mathbf{z}}}
\def\b0{{\mathbf{0}}}

\def\bA{{\mathbf{A}}}
\def\bB{{\mathbf{B}}}
\def\bC{{\mathbf{C}}}
\def\bD{{\mathbf{D}}}
\def\bE{{\mathbf{E}}}
\def\bF{{\mathbf{F}}}
\def\bG{{\mathbf{G}}}
\def\bH{{\mathbf{H}}}
\def\bI{{\mathbf{I}}}
\def\bJ{{\mathbf{J}}}
\def\bK{{\mathbf{K}}}
\def\bL{{\mathbf{L}}}
\def\bM{{\mathbf{M}}}
\def\bN{{\mathbf{N}}}
\def\bO{{\mathbf{O}}}
\def\bP{{\mathbf{P}}}
\def\bQ{{\mathbf{Q}}}
\def\bR{{\mathbf{R}}}
\def\bS{{\mathbf{S}}}
\def\bT{{\mathbf{T}}}
\def\bU{{\mathbf{U}}}
\def\bV{{\mathbf{V}}}
\def\bW{{\mathbf{W}}}
\def\bX{{\mathbf{X}}}
\def\bY{{\mathbf{Y}}}
\def\bZ{{\mathbf{Z}}}

\def\mA{{\mathbb{A}}}
\def\mB{{\mathbb{B}}}
\def\mC{{\mathbb{C}}}
\def\mD{{\mathbb{D}}}
\def\mE{{\mathbb{E}}}
\def\mF{{\mathbb{F}}}
\def\mG{{\mathbb{G}}}
\def\mH{{\mathbb{H}}}
\def\mI{{\mathbb{I}}}
\def\mJ{{\mathbb{J}}}
\def\mK{{\mathbb{K}}}
\def\mL{{\mathbb{L}}}
\def\mM{{\mathbb{M}}}
\def\mN{{\mathbb{N}}}
\def\mO{{\mathbb{O}}}
\def\mP{{\mathbb{P}}}
\def\mQ{{\mathbb{Q}}}
\def\mR{{\mathbb{R}}}
\def\mS{{\mathbb{S}}}
\def\mT{{\mathbb{T}}}
\def\mU{{\mathbb{U}}}
\def\mV{{\mathbb{V}}}
\def\mW{{\mathbb{W}}}
\def\mX{{\mathbb{X}}}
\def\mY{{\mathbb{Y}}}
\def\mZ{{\mathbb{Z}}}

\def\cA{\mathcal{A}}
\def\cB{\mathcal{B}}
\def\cC{\mathcal{C}}
\def\cD{\mathcal{D}}
\def\cE{\mathcal{E}}
\def\cF{\mathcal{F}}
\def\cG{\mathcal{G}}
\def\cH{\mathcal{H}}
\def\cI{\mathcal{I}}
\def\cJ{\mathcal{J}}
\def\cK{\mathcal{K}}
\def\cL{\mathcal{L}}
\def\cM{\mathcal{M}}
\def\cN{\mathcal{N}}
\def\cO{\mathcal{O}}
\def\cP{\mathcal{P}}
\def\cQ{\mathcal{Q}}
\def\cR{\mathcal{R}}
\def\cS{\mathcal{S}}
\def\cT{\mathcal{T}}
\def\cU{\mathcal{U}}
\def\cV{\mathcal{V}}
\def\cW{\mathcal{W}}
\def\cX{\mathcal{X}}
\def\cY{\mathcal{Y}}
\def\cZ{\mathcal{Z}}
\def\cd{\mathcal{d}}
\def\Mt{M_{t}}
\def\Mr{M_{r}}
\def\O{\Omega_{M_{t}}}
\newcommand{\figref}[1]{{Fig.}~\ref{#1}}
\newcommand{\tabref}[1]{{Table}~\ref{#1}}

\newcommand{\var}{\mathsf{var}}
\newcommand{\fb}{\tx{fb}}
\newcommand{\nf}{\tx{nf}}
\newcommand{\BC}{\tx{(bc)}}
\newcommand{\MAC}{\tx{(mac)}}
\newcommand{\Pout}{p_{\mathsf{out}}}
\newcommand{\nnn}{\nn\\}
\newcommand{\FB}{\tx{FB}}
\newcommand{\TX}{\tx{TX}}
\newcommand{\RX}{\tx{RX}}
\renewcommand{\mod}{\tx{mod}}
\newcommand{\m}[1]{\mathbf{#1}}
\newcommand{\td}[1]{\tilde{#1}}
\newcommand{\sbf}[1]{\scriptsize{\textbf{#1}}}
\newcommand{\stxt}[1]{\scriptsize{\textrm{#1}}}
\newcommand{\suml}[2]{\sum\limits_{#1}^{#2}}
\newcommand{\sumlk}{\sum\limits_{k=0}^{K-1}}
\newcommand{\eqhsp}{\hspace{10 pt}}
\newcommand{\tx}[1]{\texttt{#1}}
\newcommand{\Hz}{\ \tx{Hz}}
\newcommand{\sinc}{\tx{sinc}}
\newcommand{\tr}{\mathrm{tr}}
\newcommand{\MAI}{\tx{MAI}}
\newcommand{\ISI}{\tx{ISI}}
\newcommand{\IBI}{\tx{IBI}}
\newcommand{\CN}{\tx{CN}}
\newcommand{\CP}{\tx{CP}}
\newcommand{\ZP}{\tx{ZP}}
\newcommand{\ZF}{\tx{ZF}}
\newcommand{\SP}{\tx{SP}}
\newcommand{\MMSE}{\tx{MMSE}}
\newcommand{\MINF}{\tx{MINF}}
\newcommand{\RC}{\tx{MP}}
\newcommand{\MBER}{\tx{MBER}}
\newcommand{\MSNR}{\tx{MSNR}}
\newcommand{\MCAP}{\tx{MCAP}}
\newcommand{\vol}{\tx{vol}}
\newcommand{\ah}{\hat{g}}
\newcommand{\tg}{\tilde{g}}
\newcommand{\teta}{\tilde{\eta}}
\newcommand{\heta}{\hat{\eta}}
\newcommand{\uh}{\m{\hat{s}}}
\newcommand{\eh}{\m{\hat{\eta}}}
\newcommand{\hv}{\m{h}}
\newcommand{\hh}{\m{\hat{h}}}
\newcommand{\Po}{P_{\mathrm{out}}}
\newcommand{\Poh}{\hat{P}_{\mathrm{out}}}
\newcommand{\Ph}{\hat{\gamma}}
\newcommand{\mat}[1]{\begin{matrix}#1\end{matrix}}
\newcommand{\ud}{^{\dagger}}
\newcommand{\C}{\mathcal{C}}
\newcommand{\nn}{\nonumber}
\newcommand{\nInf}{U\rightarrow \infty}
\clearpage{}

\title{Generative Deep Learning for CSI Feedback in Massive MIMO via Channel Distribution Prediction
}

\author{Liujia Yao, Changsheng You, Chao Zhou, Yi Gong, and Chengwen Xing\thanks{L. Yao, C. You, C. Zhou, and Y. Gong are with the Department of Electronic and Electrical Engineering, Southern University of Science and Technology, Shenzhen 518055, China (e-mail: \{yaolj2024, zhouchao2024\}@mail.sustech.edu.cn; \{youcs, gongy\}@sustech.edu.cn). C. Xing is with the School of Information and Electronics, Beijing Institute of Technology, Beijing 100081, China (e-mail: chengwenxing@bit.edu.cn).
The source code that reproduces the main results of this work will be made publicly available at \url{https://github.com/ensomnia16/PAC} upon acceptance.
}
}
\maketitle

\begin{abstract}
Existing deep-learning-based channel state information (CSI) feedback methods typically reduce feedback overhead by compressing high-dimensional CSI into low-dimensional representations
and/or exploit historical CSI to predict future CSI.
However, these methods may become \emph{unreliable} when user equipment (UE) suddenly changes its motion.
To address this issue, we propose in this paper a new \emph{prediction-aided coding} (PAC) framework for generative CSI feedback, which adaptively encodes CSI for feedback based on previously predicted channel/CSI distributions and the observed one in real time.
In particular, each CSI frame is represented by multiple CSI tokens obtained without using historical CSI, which renders CSI token generation \emph{robust} to abrupt UE motion.
Based on this CSI token representation,
in the first stage, the \emph{conditional distribution} (instead of a deterministic value) of the next-frame CSI tokens is predicted based on historical ones.
Then, in the second stage, once the next frame arrives,
such a predicted distribution is exploited for \emph{variable-length} entropy coding of the CSI tokens.
To further reduce the feedback overhead and inference latency of PAC, we propose efficient designs for the two stages.
Specifically, for the first stage, a customized causal Transformer with a new positional embedding strategy is designed to accurately predict the probabilities of next-frame CSI tokens.
While for the second stage,
to reduce the inference latency caused by sequential token decoding, we develop a \emph{parallel} causal decoding strategy.
Finally, extensive experiments show that compared with existing history-aware baselines, the proposed PAC reduces feedback overhead by approximately $72\%$ to $84\%$, while achieving higher reconstruction accuracy.
Additionally, PAC maintains CSI reconstruction accuracy under random-walk trajectories, whereas existing history-aware baselines suffer sharp degradation in CSI reconstruction accuracy.
\end{abstract}
\begin{IEEEkeywords}
CSI feedback, massive MIMO, deep learning, generative AI, causal Transformer.
\end{IEEEkeywords}

\vspace{-0.2em}
\section{Introduction}

Massive multiple-input multiple-output (MIMO) has been widely recognized
as a key technology for next-generation wireless communication systems,
owing to its capability to substantially improve spectral efficiency and link reliability, while recent advances further extend it to near-field and far-field communication and sensing
~\cite{You2025NGAT,lee2009mimo,zhang2023mixedfield,zhou2026beamcoverage,liu2026mixedfield,liu2023nearfieldtutorial,li2026localization}.
To fully exploit
these advantages, base stations (BSs) require accurate downlink channel
state information (CSI) for designing efficient beamforming~\cite{wen2018csinet,yao2026codebook}.
In time division duplex (TDD) systems, BSs can acquire the downlink CSI from uplink pilots through channel reciprocity using advanced estimation schemes, such as polar-domain and learning-based methods~\cite{cui2022nearfield,gao2024xlcnet}.
In frequency division duplex
(FDD) systems, however, the downlink CSI cannot be directly obtained
from uplink channel estimation due to uplink-downlink non-reciprocity, thus necessitating efficient CSI feedback methods.
Moreover, as the numbers of antennas and sub-carriers
increase, the downlink CSI matrix becomes increasingly high-dimensional; hence, directly feeding back the full CSI matrix will inevitably incur prohibitively high CSI feedback overhead~\cite{wen2018csinet,liu2022markovnet}.

Although history-aware deep learning (DL)-based methods have been developed for CSI feedback and reconstruction, their accuracy degrades significantly when user equipment (UE) experiences abrupt motion changes, such as sudden turns or stops.
To address this issue, we propose in this paper a new \emph{prediction-aided coding} (PAC) framework for \emph{generative} CSI feedback,
in which the \emph{conditional distribution} (instead of a deterministic value) of CSI tokens is predicted based on historical ones.
Such a predicted distribution is further exploited for variable-length entropy coding of CSI tokens to achieve reliable and low-overhead CSI feedback.
Thanks to the lossless nature of entropy coding, CSI tokens can be losslessly reconstructed/generated at the BS even when the predicted CSI distribution is inaccurate.

\vspace{-0.2em}
\subsection{Related Works}
Directly feeding back the full CSI matrix incurs prohibitively high CSI feedback overhead, particularly in massive MIMO systems with large-scale antenna arrays.
To reduce the CSI feedback dimensionality, conventional compressed-sensing-based methods exploit the sparse representation of high-dimensional CSI matrices over predefined codebooks, such as the discrete Fourier transform (DFT)-based angular-delay codebook~\cite{zhang2023mixedfield,wu2024dftbeamtraining} and/or the polar-domain codebook~\cite{zhangFastBT}.
Based on this sparse representation, the UE compresses its CSI into a low-dimensional measurement vector, which is then quantized and fed back to the BS for CSI reconstruction~\cite{wen2018csinet,guo2022overview}.
However, the effectiveness of these methods depends heavily on how well the predefined codebook matches the underlying channel structure.
Since channel characteristics generally vary across dynamic propagation environments, a predefined codebook may suffer from channel model mismatch, resulting in an \emph{inaccurate channel representation} and degraded CSI reconstruction accuracy.
Moreover, as the number of antennas increases, the CSI dimension and the number of measurements required for reliable CSI reconstruction also increase, which incurs \emph{considerable} CSI feedback overhead.

To overcome the limitations of conventional CSI feedback methods, several DL-based CSI feedback methods were developed (see, e.g., \cite{wen2018csinet,guo2020csinetplus,lu2020crnet,deepcmc2021,liu2022markovnet,wang2019csinetlstm,lu2019reccsinet,liu2019dualnet}), in which each CSI frame is treated as an independent ``image'', and neural networks are employed to compress it into a compact representation for efficient CSI feedback.
As a representative DL-based CSI feedback method, CsiNet~\cite{wen2018csinet} employed a convolutional neural network (CNN)-based autoencoder to compress the CSI matrix into a low-dimensional latent vector, thereby achieving a high compression ratio with accurate CSI reconstruction.
Alternatively,
DeepCMC~\cite{deepcmc2021} designed a CNN-based autoencoder, in which the entropy of the scalar-quantized latent representation was jointly trained with the CSI reconstruction errors to strike a balance between CSI feedback overhead and reconstruction accuracy.
Furthermore, attention mechanisms were introduced in~\cite{cui2022transnet} to further improve CSI reconstruction accuracy,
which effectively exploit the spatial correlation across different antennas and sub-carriers that is difficult to capture by using CNNs alone.
More recently,
vector quantization (VQ)-based methods were proposed in \cite{vqvae} to map the continuous latent vector into a sequence of discrete codeword indices selected from a codebook, such that only the indices, rather than the full-precision latent coefficients, need to be fed back.
To further reduce CSI feedback overhead, an entropy-constrained vector-quantized variational autoencoder (ECVQ-VAE)-based method was proposed in~\cite{shin2025ecvqvae}.
It promotes a low-entropy distribution over the learned codeword indices, such that frequently selected indices can be encoded with shorter codewords, thereby greatly reducing average CSI feedback overhead.
Nevertheless, all the above methods are \emph{history-agnostic}, which generally lack the capability to exploit the temporal correlation across CSI frames for fast CSI feedback.

Subsequently, a set of history-aware DL-based CSI feedback methods were recently proposed in the literature (see, e.g.,~\cite{wang2019csinetlstm,li2020spatiotemporal,liu2022markovnet,zhang2025futurecsi,compact}).
These methods exploit temporal correlation across consecutive CSI frames to predict the next CSI frame,
thereby reducing the amount of information needed to be fed back.
For example, the authors in \cite{wang2019csinetlstm} proposed a history-aware CSI feedback method in which the UE uses the CsiNet encoder to compress the current CSI into a low-dimensional representation for feedback.
At the BS, a long short-term memory (LSTM) module exploits historical CSI to refine the reconstruction of current CSI from the received compressed representation, thereby achieving more accurate CSI reconstruction than history-agnostic CsiNet.
To reduce the model complexity of neural networks, the authors in~\cite{liu2022markovnet} adopted a first-order Markov predictor,
where the UE encodes and feeds back the difference between current CSI and its Markovian prediction only~\cite{liu2022markovnet}, thereby reducing both CSI feedback overhead and the number of model parameters.
More recently, attention-based architectures were introduced in~\cite{compact} to model long-range temporal dependency
that may be difficult to capture by using recurrent or Markovian models.
For instance, the authors in~\cite{compact} exploited the sparsity of millimeter-wave (mmWave) channels and represented each CSI frame
by using relatively few path parameters (e.g., angles, delays, and channel gains).
Then, a Transformer was used to predict path parameters from their historical observations at the UE, which were then quantized and fed back to the BS.
Although these methods can effectively reduce CSI feedback overhead,
their CSI reconstruction accuracy critically hinges on the prediction accuracy of current CSI from historical CSI.
When a UE undergoes an abrupt motion change, such as a sudden turn or stop~\cite{3gpp_tr36873}, the CSI prediction accuracy can deteriorate significantly, resulting in \emph{severe degradation} in CSI reconstruction accuracy.
Therefore, how to exploit temporal correlation to reduce CSI feedback overhead while maintaining satisfactory CSI reconstruction accuracy, particularly under abrupt UE mobility changes and/or out-of-distribution motion patterns, remains an open problem.

\vspace{-0.2em}
\subsection{Contributions}
Motivated by the above, this paper aims to overcome the aforementioned limitations of existing history-agnostic and history-aware CSI feedback methods.
Specifically, we consider an FDD massive MIMO system, in which a UE generally moves along a smooth trajectory (e.g., a straight trajectory) and periodically feeds back its downlink CSI to the BS.
However, the UE may unexpectedly change its motion, for example, by making a sudden turn or coming to a stop because of unforeseen events.
To reduce CSI feedback overhead while maintaining accurate CSI reconstruction under such abrupt channel changes,
we propose a new history-aware PAC framework for fast and accurate CSI feedback by exploiting neural compression methods \cite{balle2017end,qian2022entroformer}.
The main contributions are summarized as follows.
\begin{itemize}
\item First, we propose PAC, a new generative CSI feedback framework for massive MIMO systems,
which comprises mainly two stages.
In the first stage, PAC predicts the conditional distribution (instead of a deterministic value) of next-frame CSI tokens based on historical CSI tokens.
In the second stage, once the next CSI frame arrives,
this predicted distribution is exploited for entropy coding of these CSI tokens into a variable-length bitstream for CSI feedback.
Note that, unlike existing history-aware methods, in which the predicted \emph{CSI} directly affects the reconstructed CSI,
PAC uses the predicted channel/CSI \emph{distribution} to determine the bitstream length in lossless entropy coding only.
Thus, inaccurate probability prediction only increases the required bitstream length for CSI feedback, yet without introducing additional CSI reconstruction error.

\item Second, we propose efficient designs for the two stages of PAC.
Specifically, we adopt an ECVQ-VAE-based tokenizer to map each CSI frame into a sequence of discrete CSI tokens.
For Stage 1,
we design a customized causal Transformer for accurate token-probability prediction.
Note that conventional Transformers use absolute positional embedding (APE) to capture the positional information of tokens in the token sequence, which ignores the intra-frame positional information of CSI tokens.
To address this issue, we propose an \emph{intra-frame positional embedding} (IPE) strategy, which enables the Transformer to capture intra-CSI-frame positional information of the CSI tokens, thereby achieving more accurate token-probability prediction.
Moreover, for Stage 2, to reduce the inference latency caused by sequential CSI token decoding in conventional Transformers, we further develop a \emph{parallel} causal decoding architecture that enables the BS to decode CSI tokens in parallel, thereby significantly reducing inference latency with mild performance loss.

\item Last, we present extensive numerical results to validate the feedback efficiency and robustness of the proposed PAC using the DeepMIMO dataset~\cite{Alkhateeb2019DeepMIMO} across diverse mobility patterns and trajectories, together with additional indoor-channel validation based on the COST 2100 dataset~\cite{cost2100}.
It is observed that the proposed PAC reduces CSI feedback overhead by approximately 72\% to 84\% compared with two representative history-aware baselines, while achieving a lower normalized mean squared error (NMSE) for CSI reconstruction.
Moreover, even when abrupt motion changes cause a history-aware baseline to experience an NMSE degradation of more than 10 decibels (dB), the proposed PAC still remains effective with nearly unchanged CSI reconstruction NMSE.
\end{itemize}

\textit{Organization:} The remainder of this paper is organized as follows. Section~\ref{sec:system} presents the system model and the problem formulation. Section~\ref{sec:framework} reviews existing benchmark schemes and introduces the proposed PAC framework. Section~\ref{sec:tokenizer} presents the ECVQ-VAE-based CSI representation. Sections~\ref{sec:transformer} and~\ref{sec:entropy_coding} present CSI distribution prediction and prediction-aided CSI entropy coding, respectively.
Section~\ref{sec:experiments} reports numerical results, followed by conclusions summarized in Section~\ref{sec:conclusion}.
The key symbols used throughout the paper are summarized in Table~\ref{tab:notation}.
\begin{table}[t]
\centering
\caption{Summary of main symbols.}
\label{tab:notation}
\renewcommand{\arraystretch}{1.15}
\setlength{\tabcolsep}{4pt}
\begin{tabular}{p{0.18\columnwidth} p{0.74\columnwidth}}
\toprule
\textbf{Symbol} & \textbf{Description} \\
\midrule
$M_{\rm tx}$, $N_{\rm c}$, $N_{\rm d}$ & Numbers of BS antennas, sub-carriers, and retained delay taps \\
$\mathbf{H}_t$, $\hat{\mathbf{H}}_t$ & Ground-truth truncated angular-delay CSI matrix and reconstructed CSI matrix at the BS \\
\midrule
$\mathbf{Z}_t$, $\mathbf{z}_{t,n}$ & Continuous latent representation and its $n$-th subvector \\
$N$, $D$ & Number of tokens per frame and token feature dimension \\
$\mathcal{C}$, $K$, $\mathbf{c}_k$ & VQ codebook, its size, and the $k$-th codeword \\
$\mathcal{K}_t$, $k_{t,n}$ & Discrete token sequence and the $n$-th token index \\
$P_k$ & Prior probability of selecting the $k$-th codeword \\
$\lambda$ & CSI feedback overhead-distortion multiplier \\
\midrule
$L$ & {Number of available CSI frames in the context window} \\
$d_{\rm m}$, $N_{\rm layer}$ & Transformer model dimension and number of Transformer layers \\
$G$, $\tilde{N}$, $\mathcal{G}_g$ & Number of parallel groups, group size, and the $g$-th group of token positions \\
{$\mathbf{p}_{t+1,n}$} & Token probabilities of token $n$ for frame $(t+1)$ \\
{$\mathcal{U}_{t+1}$} & Arithmetic-coding mapping for frame $(t+1)$ \\
{$\mathbf{b}_{t+1}$} & {CSI feedback bitstream for frame $(t+1)$} \\
$\{0,1\}^*$ & Set of all finite-length binary strings \\
\bottomrule
\end{tabular}
\vspace{-1.8em}
\end{table}

\vspace{-0.2em}
\section{System Model and Problem Statement}\label{sec:system}
We consider a downlink massive MIMO-orthogonal frequency division multiplexing (OFDM) system, in which a BS equipped with $M_{\rm tx} \gg 1$ antennas serves multiple single-antenna UEs over $N_{\rm c}$ sub-carriers.

\vspace{-0.2em}
\subsection{MIMO-OFDM Signal and Channel Models}

For a typical UE, the received signal over the $n_{\rm c}$-th sub-carrier at time slot $t$ is given by
\begin{equation}
y_{n_{\rm c},t} = \tilde{\mathbf{h}}_{n_{\rm c},t}^H \mathbf{w}_{n_{\rm c},t} x_{n_{\rm c},t} + \eta_{n_{\rm c},t}, \forall n_{\rm c} \in \{1,\dots,N_{\rm c}\},
\end{equation}
where $\tilde{\mathbf{h}}_{n_{\rm c},t} \in \mathbb{C}^{M_{\rm tx} \times 1}$ is the downlink channel vector, $\mathbf{w}_{n_{\rm c},t} \in \mathbb{C}^{M_{\rm tx} \times 1}$ is the precoding vector, $x_{n_{\rm c},t}$ is the transmitted symbol, and $\eta_{n_{\rm c},t} \sim \mathcal{CN}(0,\sigma^2)$ represents additive white Gaussian noise (AWGN).
By stacking the channel vectors across all $N_{\rm c}$ sub-carriers, we define the spatial-frequency channel matrix at time $t$ as $\tilde{\mathbf{H}}_t = [\tilde{\mathbf{h}}_{1,t}, \dots, \tilde{\mathbf{h}}_{N_{\rm c},t}]^H \in \mathbb{C}^{N_{\rm c} \times M_{\rm tx}}$.

In massive MIMO-OFDM systems, the channel responses are highly correlated across different sub-carriers and antennas.
Thus, the spatial-frequency channel matrix $\tilde{\mathbf{H}}_t$ exhibits sparsity in the angular-delay domain.
Specifically, by performing a two-dimensional (2D) DFT along the spatial and frequency domains, the angular-delay domain channel representation, denoted by $\mathbf{H}_{t}^{\rm ad}$, is given by
\begin{equation}
\mathbf{H}_{t}^{\rm ad} = \mathbf{F}_{\rm d} \tilde{\mathbf{H}}_t \mathbf{F}_{\rm a}^H,
\end{equation}
where $\mathbf{F}_{\rm d} \in \mathbb{C}^{N_{\rm c} \times N_{\rm c}}$ and $\mathbf{F}_{\rm a} \in \mathbb{C}^{M_{\rm tx} \times M_{\rm tx}}$ denote the DFT matrices for the delay-domain and angular-domain transformations, respectively.
Mathematically, for a $Q$-dimensional DFT matrix $\mathbf{F}_Q$, its $(p, q)$-th entry is given by $[\mathbf{F}_Q]_{p,q} = \frac{1}{\sqrt{Q}} e^{-\jmath \frac{2\pi}{Q}(p-1)(q-1)}, \forall p,q \in \{1,\dots,Q\}$.
Since the channel matrix is generally sparse in the angular-delay domain, most of the channel power is concentrated on a small number of dominant angular-delay components~\cite{wen2018csinet}.
As such, the angular-delay domain channel matrix $\mathbf{H}_{t}^{\rm ad}$ can be truncated by retaining only the first $N_{\rm d}$ rows ($N_{\rm d} \ll N_{\rm c}$), yielding a reduced-dimensional effective channel matrix as follows
\begin{equation}
\mathbf{H}_t =
\begin{bmatrix}
h_{1,1,t} & \dots & h_{1,M_{\rm tx},t} \\
\vdots & \ddots & \vdots \\
h_{N_{\rm d},1,t} & \dots & h_{N_{\rm d},M_{\rm tx},t}
\end{bmatrix}
\in \mathbb{C}^{N_{\rm d} \times M_{\rm tx}}.
\end{equation}
Herein, $h_{d,m,t} \in \mathbb{C}, \forall d\in\{1,\dots,N_{\rm d}\}, m\in\{1,\dots,M_{\rm tx}\}$ is the channel coefficient at the {$d$-th} delay tap and the $m$-th angular direction at time slot $t$.
Note that the truncated CSI $\mathbf{H}_t$ retains most of the channel information and is therefore used as the \emph{reference} for CSI feedback, avoiding the excessive overhead of feeding back the full angular-delay-domain channel $\mathbf{H}_{t}^{\rm ad}$.
\vspace{-0.2em}
\subsection{CSI Feedback Problem and Design Challenge}
To design transmit beamforming for downlink data transmission, the BS requires accurate downlink CSI from the UE.
Specifically, at each time slot, the UE encodes the truncated CSI $\mathbf{H}$ into a finite-length bitstream $\mathbf{b} = \mathcal{E}(\mathbf{H}) \in \{0,1\}^*$ and sends it to the BS\footnote{
In 5G NR, CSI is fed back over the physical uplink control channel (PUCCH).
The PUCCH uses strong polar coding~\cite{3gpp_ts38212} together with hybrid automatic repeat request retransmission, making bit errors negligible in practice, especially for the low-feedback-overhead CSI feedback considered in this work.
Thus, we assume an error-free uplink CSI feedback channel, which has been widely adopted in the existing literature~\cite{wen2018csinet,liu2022markovnet,wang2019csinetlstm}.}.
Based on this bitstream, the BS reconstructs CSI as $\hat{\mathbf{H}} = \mathcal{D}(\mathbf{b})$, where the mappings $\mathcal{E}(\cdot)$ and $\mathcal{D}(\cdot)$ denote the CSI feedback encoder at the UE and the decoder at the BS, respectively.
The feedback overhead is measured by the required codeword length $|\mathbf{b}|$ in bits, whereas the CSI reconstruction accuracy is measured by CSI reconstruction NMSE, defined as $\mathbb{E}\!\left[ \|\mathbf{H} - \hat{\mathbf{H}}\|_F^2 / \|\mathbf{H}\|_F^2 \right]$.
Note that, even after channel matrix truncation, directly feeding back the raw parameters of $\mathbf{H}$ at every time slot would still incur prohibitively high CSI feedback overhead due to the high dimensionality of $\mathbf{H}$ and the limited CSI feedback bandwidth.
For example, consider a CSI tensor of size $32\times 32 \times 2$, with $N_{\rm d}=32$, $M_{\rm tx}=32$, and the real and imaginary parts represented by two separate channels.
When each real-valued entry is quantized with 16 bits, the total feedback overhead is up to $32 \times 32 \times 2 \times 16 = 32768$ bits per frame (bpf), which is unaffordable for real-time CSI feedback in practical systems.
Therefore, how to reduce the feedback overhead while maintaining CSI reconstruction accuracy remains a critical challenge in CSI feedback design.

\vspace{-0.2em}
\section{Benchmark Schemes and Proposed CSI Feedback Framework}\label{sec:framework}
In this section, we first review two representative benchmark schemes for DL-based CSI feedback and discuss their limitations.
Then, we propose a new history-aware CSI feedback framework to overcome these limitations.
\vspace{-0.2em}
\subsection{Benchmark Schemes}
\subsubsection{History-Agnostic CSI Feedback}
History-agnostic CSI compression methods employ a DL-based encoder to map the high-dimensional CSI matrix into a low-dimensional latent representation.
For example, by using a vector-quantized variational autoencoder (VQ-VAE) architecture, the continuous CSI matrix can be quantized into a sequence of discrete tokens drawn from a learned codebook~\cite{vqvae}, which can reduce CSI feedback overhead.
However, such history-agnostic CSI compression methods fail to exploit the strong temporal correlation across consecutive CSI frames, which restricts their capability to further reduce CSI feedback overhead.

\subsubsection{History-Aware CSI Feedback}
History-aware CSI feedback methods exploit temporal correlation among consecutive CSI frames to reduce CSI feedback overhead.
Existing history-aware CSI feedback methods fall into two representative categories~\cite{guo2022overview}.
The first category refers to recurrent-neural-network-based methods~\cite{wang2019csinetlstm}, in which a UE first compresses the CSI into a low-dimensional representation via CsiNet and then feeds it back to the BS.
At the BS, an LSTM module is used to refine the reconstruction of the CSI from the received compressed representation by exploiting historical CSI.
The second category refers to Markovian differential CSI feedback methods, in which the UE encodes and feeds back only the residual/discrepancy between the ground-truth CSI and the model-predicted CSI.
However, although these methods perform well under smooth UE motion, their CSI prediction becomes \emph{unreliable} in out-of-distribution scenarios involving abrupt motion changes, such as sudden turns and stops~\cite{3gpp_tr36873}, resulting in significant degradation in CSI reconstruction accuracy.

\begin{figure}[t]
\centering
\includegraphics[width=\linewidth]{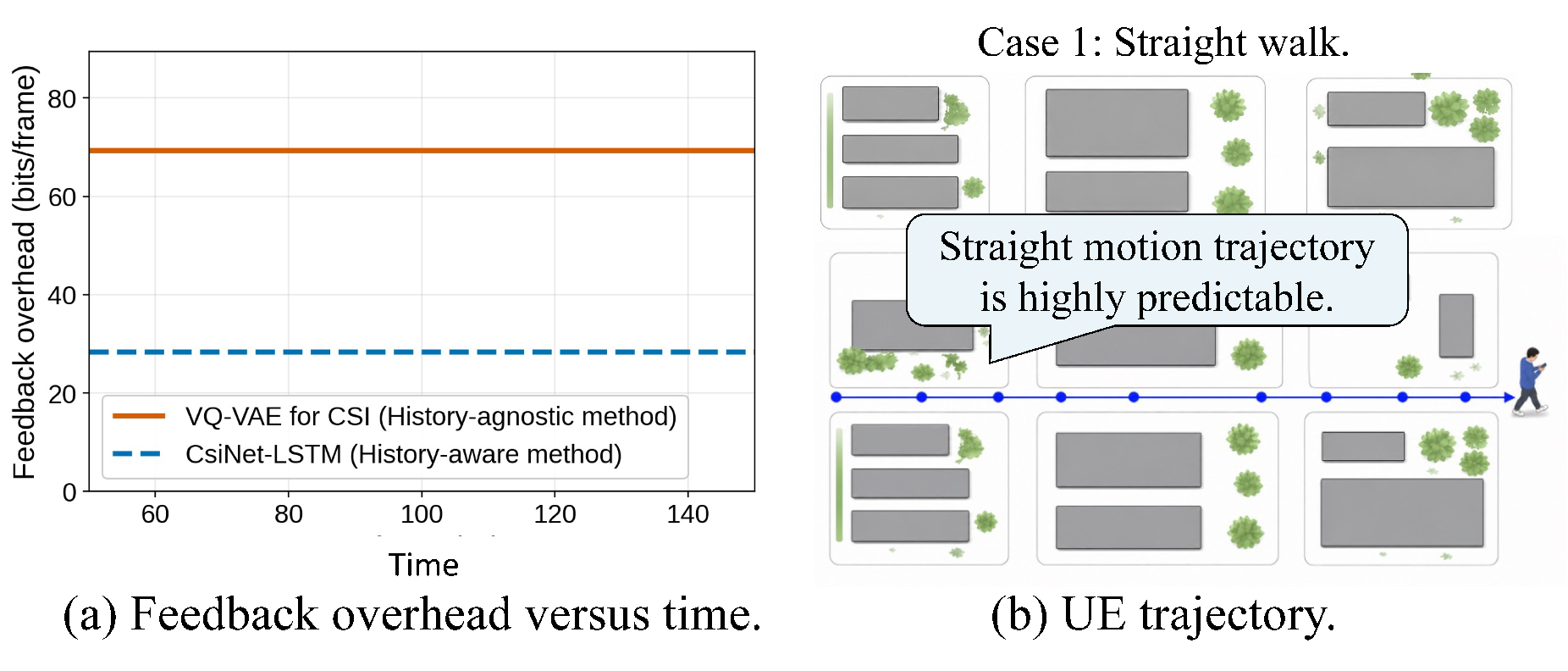}
\caption{Feedback overhead of the history-agnostic method and the history-aware method in Case 1, both achieving a similar NMSE of approximately $-12.4$ dB.}
\label{fig:inefficient}
\vspace{-1.2em}
\end{figure} 
\begin{figure}[t]
\centering
\includegraphics[width=\linewidth]{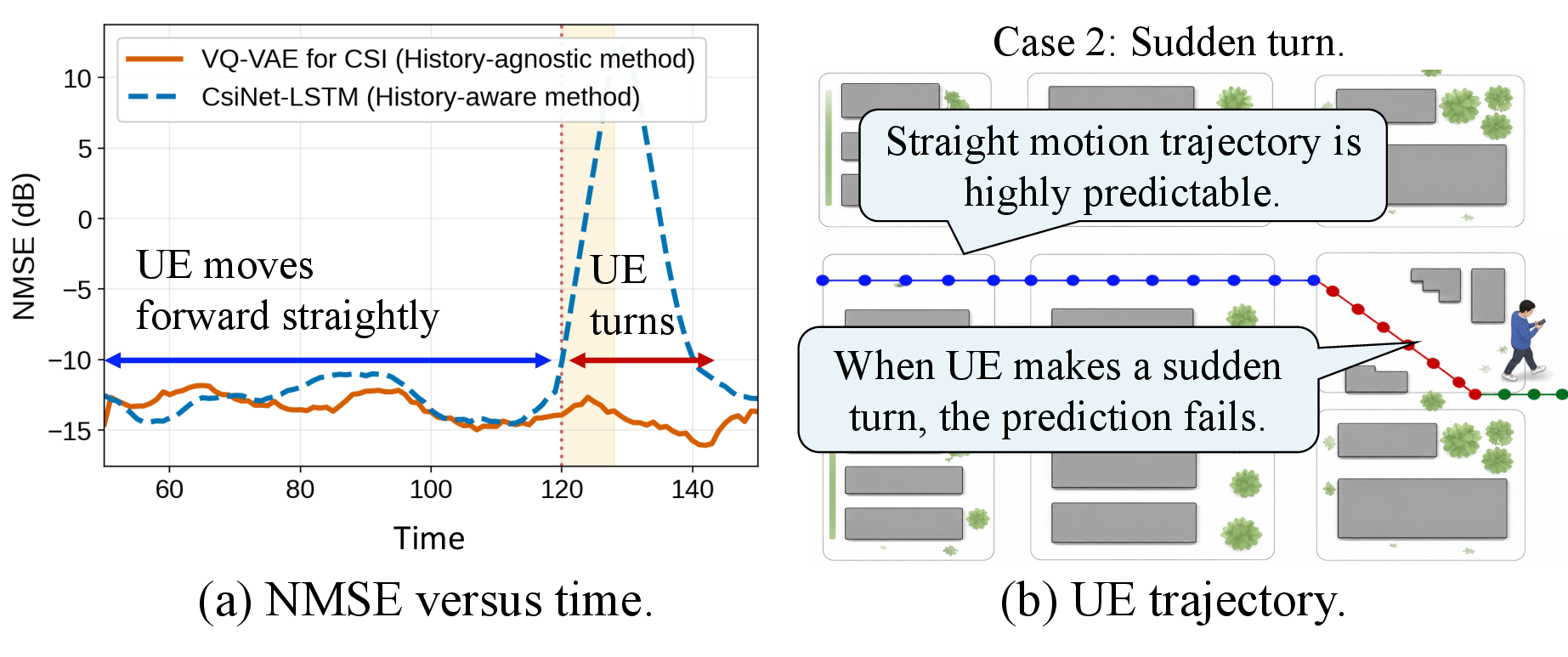}
\caption{NMSE of the history-agnostic method and the history-aware method in Case 2, where the UE makes a sudden right turn during a 9-frame window (i.e., frames $120$–$128$).}
\label{fig:fragile}
\vspace{-1.2em}
\end{figure} 

\begin{figure*}[ht]
\centering
\includegraphics[width=0.95\textwidth]{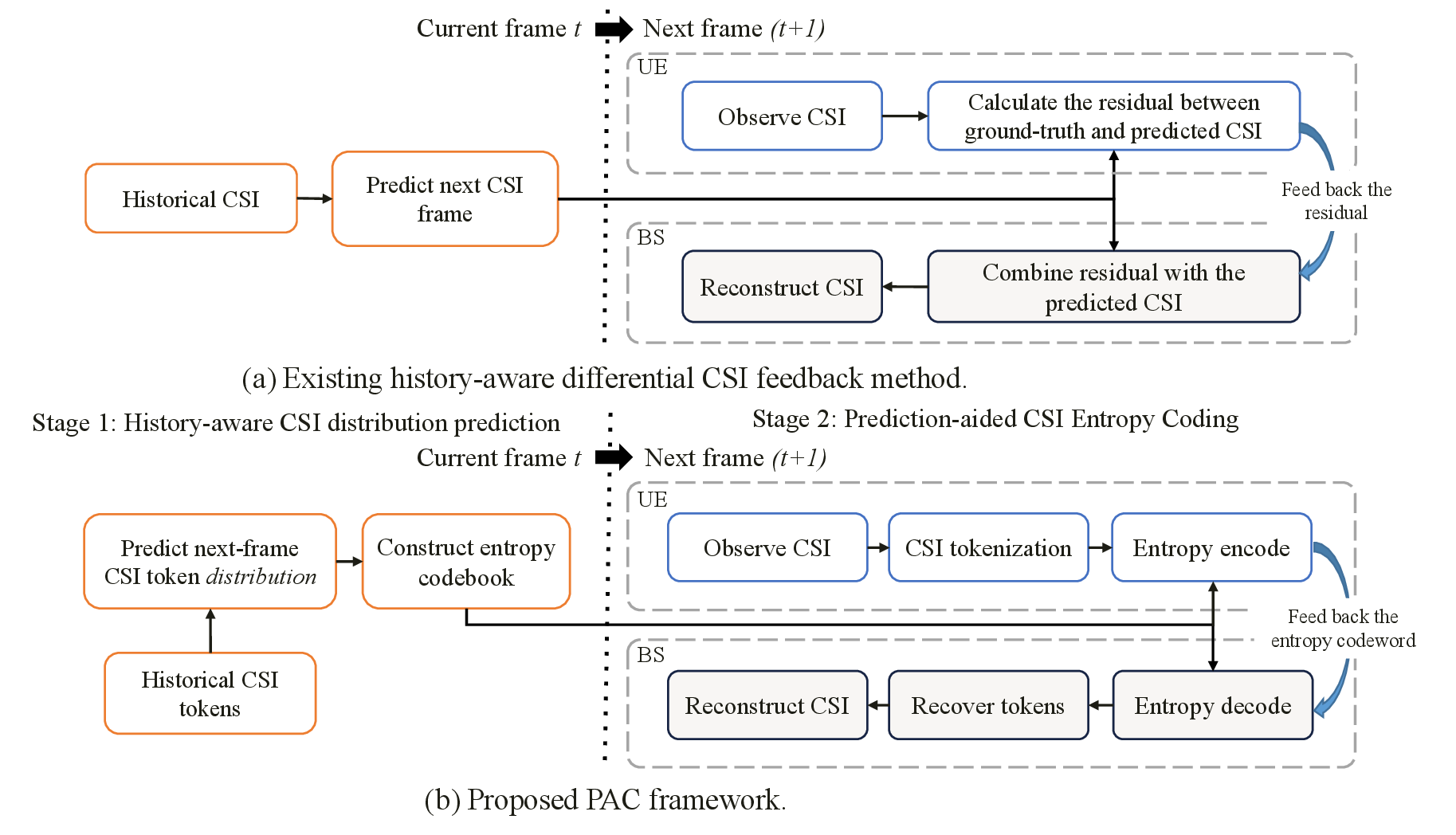}
\caption{Comparison between (a) existing history-aware differential CSI feedback, and (b) the proposed PAC, where the predicted token distribution is used for entropy coding of the CSI tokens only.}
\label{fig:pipeline}
\vspace{-1.4em}
\end{figure*}
\begin{example}[Excessive CSI feedback overhead of history-agnostic methods]
In Fig.~\ref{fig:inefficient}, we numerically evaluate the CSI feedback overhead of representative history-agnostic and history-aware CSI feedback methods when a UE moves along a straight trajectory at $2$ meters/second (m/s), which is referred to as Case 1.
The CSI matrix of each frame has a dimension of $32\times 32 \times 2$ and is compressed using the VQ-VAE-based method~\cite{vqvae},
which compresses each CSI frame into a sequence of discrete tokens and then encodes them into a fixed-length bitstream for feedback.
The history-agnostic method~\cite{vqvae} needs approximately $69.4$ bpf for feedback to achieve an NMSE of $-12.44$ dB.
It is observed that the history-aware CSI feedback method in~\cite{wang2019csinetlstm} reduces the frame-level feedback overhead to approximately $28.4$ bpf at a similar CSI reconstruction NMSE of $-12.44$ dB.
\end{example}
\begin{example}[Fragility of existing history-aware CSI feedback methods]
In Fig.~\ref{fig:fragile}, under the same outdoor setting as in Fig.~\ref{fig:inefficient}, we consider a UE that moves along a trajectory over a $101$-frame interval (i.e., frame indices $50$–$150$) and makes a sudden right turn during a $9$-frame window (i.e., frames $120$–$128$), which is referred to as Case 2.
Before the right turn, the LSTM-based history-aware CSI feedback method achieves a CSI reconstruction NMSE of approximately $-12.4$ dB because the straight trajectory is smooth and hence the CSI prediction is accurate.
However, during and immediately after the sudden turn, the NMSE degrades by more than $20$ dB (i.e., the NMSE exceeds $+10$ dB), which is caused by the inaccurate CSI prediction from historical CSI.
\end{example}

\vspace{-0.2em}
\subsection{Proposed PAC Framework}
\begin{figure}[t]
\centering
\includegraphics[width=.95\linewidth]{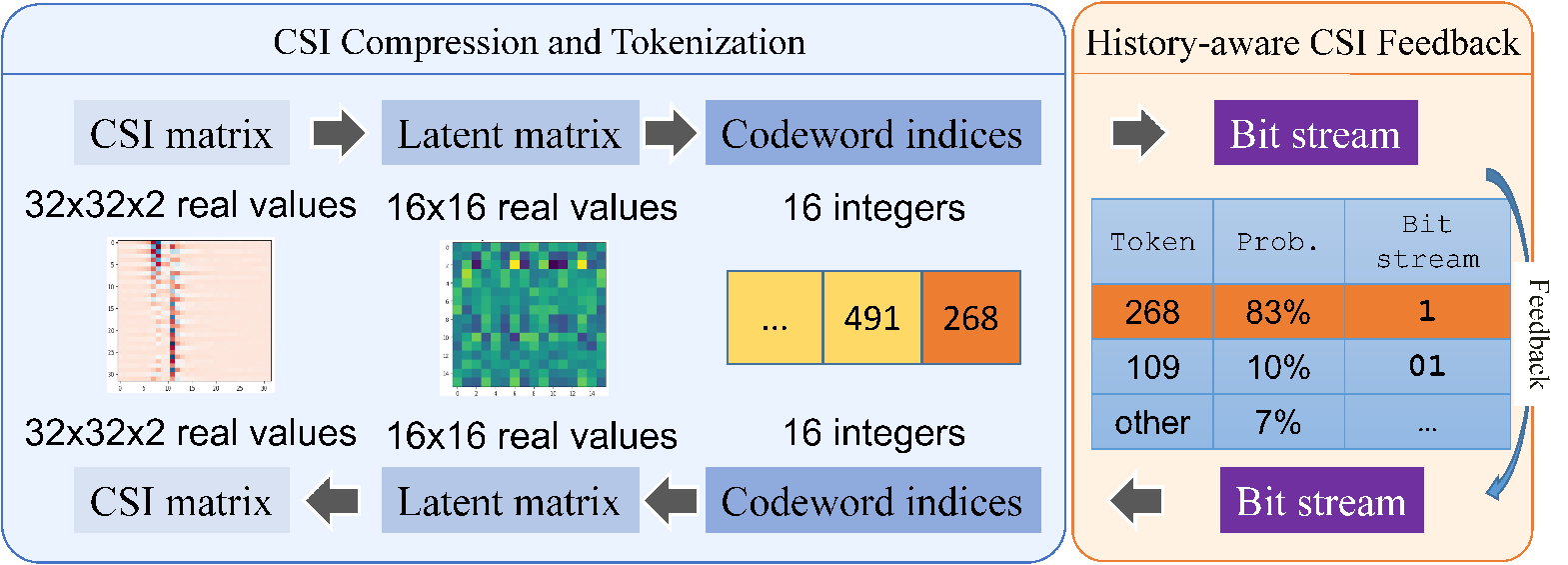}
\caption{The overall flow of the proposed CSI feedback framework, where $N_{\rm d}=M_{\rm tx}=32$, $D=N=16$, and $K=512$.}
\label{fig:transformer_flow}
\vspace{-1.5em}
\end{figure}

To address the limitations above, we propose a new history-aware CSI feedback method, called PAC, in which temporal correlation across CSI frames is exploited to predict the conditional distribution of CSI tokens based on historical CSI tokens.
Such a predicted token distribution is used to determine the length of the bitstream in lossless entropy coding only, rather than directly affecting the reconstructed CSI, as illustrated in Fig.~\ref{fig:pipeline}.  
To reduce the dimensionality of the CSI representation, 
before the distribution prediction and entropy coding,
we use an ECVQ-VAE-based tokenizer to convert each CSI frame into a sequence of $N$ discrete CSI tokens, each of which is represented by a $D$-dimensional codeword selected from a learned codebook of size $K$, as illustrated in Fig.~\ref{fig:transformer_flow}.
Note that this CSI representation does not rely on any historical CSI information,
which renders the CSI representation \emph{robust} to abrupt channel changes and out-of-distribution scenarios.
Based on the tokenized CSI representation,
the proposed PAC framework consists of two stages, namely, 1) History-aware CSI distribution prediction, and 2) Prediction-aided CSI entropy coding.
These two stages are briefly introduced below, with further details provided later.
\begin{itemize}
  \item \textbf{(Stage 1: History-aware CSI distribution prediction)}
At each current frame {$t$}, the causal Transformer uses history CSI tokens to predict the conditional distribution of next-frame CSI tokens, based on which an entropy codebook is constructed.
Specifically, the conventional Transformer uses APE to capture the positional information of tokens in the token sequence, which ignores the intra-frame positional information of CSI tokens.
To address this issue, we propose a new \emph{IPE} strategy, which enables the Transformer to capture the \emph{intra-CSI-frame positional information} of the CSI tokens, thereby achieving more accurate token-probability prediction.
\item \textbf{{(Stage 2: Prediction-aided CSI entropy coding)}}
Once the next CSI frame {$\mathbf{H}_{t+1}$} arrives, the UE obtains its CSI tokens $\mathcal{K}_{t+1}$ and encodes them using the entropy codebook constructed in Stage 1,
yielding a variable-length bitstream $\mathbf{b}_{t+1}$ for CSI feedback.
In particular, instead of the conventional Transformer that decodes CSI tokens in a serial autoregressive manner, with considerably long inference latency, we propose a \emph{parallel} causal decoding architecture,
which enables the BS to decode CSI tokens in parallel,
thereby significantly reducing the inference latency.
\end{itemize}

Since the UE and BS share the \emph{same} historical tokens and Transformer model, they can obtain the same conditional probabilities for tokens of frame $(t+1)$
without additional signaling.
Let $p_{t+1,n,k}, \forall n\in\{0,\dots,N-1\}, k\in\{0,\dots,K-1\}
$ denote the predicted probability of the $n$-th token in frame $(t+1)$ being the $k$-th VQ codeword.
Then, the distribution of the $n$-th token in frame $(t+1)$ is given by $\mathbf{p}_{t+1,n} = [p_{t+1,n,0},\dots,p_{t+1,n,K-1}]^T \in \mathbb{R}^{K\times 1}$.
Based on such predicted token distributions,
the UE and BS can construct the same entropy codebook $\mathcal{U}_{t+1}$, with the resulting bitstream length for feeding back $\mathcal{K}_{t+1}$ being approximately~\cite{cover2006elements}
\begin{equation}
\label{eq:bits_per_frame}|\mathbf{b}_{t+1}| \approx -\sum_{n=0}^{N-1}\log_2 [\mathbf{p}_{t+1,n}]_{k_{t+1,n}},
\end{equation}
where $[\mathbf{p}_{t+1,n}]_{k_{t+1,n}}$ is the probability of the ground-truth CSI token $k_{t+1,n}$. After decoding $\mathcal{K}_{t+1}$, the BS looks up the corresponding VQ codewords and uses the CNN decoder to\emph{ reconstruct/generate} $\hat{\mathbf{H}}_{t+1}$.
\vspace{-0.2em}
\section{History-Agnostic CSI Token Representation}\label{sec:tokenizer}
In this section, we employ an ECVQ-VAE to compress each CSI frame into a low-dimensional latent representation and quantize it into a sequence of discrete CSI tokens.
\begin{table}[t]
  \vspace{0.06in}
  \centering
  \caption{ECVQ-VAE encoder and decoder architectures.}
  \label{tab:ecvq_architecture}
  {\footnotesize
  \setlength{\tabcolsep}{3pt}
  \begin{tabular}{l c c}
    \toprule
    \textbf{Layer} & \textbf{Output Tensor Shape} & \textbf{Configuration} \\
    \midrule
    \multicolumn{3}{c}{\textit{Encoder} (Input: $2 \times N_{\rm d} \times M_{\rm tx}$)} \\
    \midrule
    \makecell[l]{Conv Block} & $32 \times (N_{\rm d}/2) \times (M_{\rm tx}/2)$ & 4$\times$4, $s=2$ \\
    \makecell[l]{Conv Block} & $64 \times (N_{\rm d}/4) \times (M_{\rm tx}/4)$ & 4$\times$4, $s=2$ \\
    \makecell[l]{Conv Block} & $128 \times (N_{\rm d}/4) \times (M_{\rm tx}/4)$ & 3$\times$3, $s=1$ \\
    \makecell[l]{Conv Block} & $128 \times (N_{\rm d}/4) \times (M_{\rm tx}/4)$ & 3$\times$3, $s=1$ \\
    2D Conv & $16 \times (N_{\rm d}/4) \times (M_{\rm tx}/4)$ & 3$\times$3, $s=1$ \\
    \makecell[l]{Linear Projection} & $N \times D$ & $-$ \\
    \midrule
    \multicolumn{3}{c}{\textit{Decoder} (Input: $N \times D$)} \\
    \midrule
    \makecell[l]{Inverse Linear Projection} & $16 \times (N_{\rm d}/4) \times (M_{\rm tx}/4)$ & $-$ \\
    \makecell[l]{Conv Block} & $128 \times (N_{\rm d}/4) \times (M_{\rm tx}/4)$ & 3$\times$3, $s=1$ \\
    \makecell[l]{Conv Block} & $128 \times (N_{\rm d}/4) \times (M_{\rm tx}/4)$ & 3$\times$3, $s=1$ \\
    \makecell[l]{Transposed Conv Block} & $64 \times (N_{\rm d}/2) \times (M_{\rm tx}/2)$ & 4$\times$4, $s=2$ \\
    \makecell[l]{Transposed Conv Block} & $32 \times N_{\rm d} \times M_{\rm tx}$ & 4$\times$4, $s=2$ \\
    2D Conv & $2 \times N_{\rm d} \times M_{\rm tx}$ & 3$\times$3, $s=1$ \\
    \bottomrule
  \end{tabular}
  }

  \vspace{0.5em}
  \parbox{\linewidth}{\footnotesize\raggedright
  A (transposed) convolutional (conv) block consists of a 2D (transposed) conv, batch normalization, and a leaky rectified linear unit activation.
  }
  \vspace{-2em}
\end{table}
\vspace{-0.2em}
\subsection{ECVQ-VAE-Based CSI Compression}
In particular, the ECVQ-VAE architecture~\cite{shin2025ecvqvae} is employed for CSI compression and tokenization,
which converts the CSI matrix of continuous values into a sequence of integers (i.e., CSI tokens), corresponding to the indices of the codewords in a learnable VQ codebook (see details below).
The ECVQ-VAE architecture consists of a customized CNN-based encoder, a VQ layer, and a customized CNN-based decoder, whose detailed configurations are summarized in Table~\ref{tab:ecvq_architecture}.

\textbf{CSI compression:}
The network input, i.e., the complex-valued CSI matrix $\mathbf{H}_t \in \mathbb{C}^{N_{\rm d} \times M_{\rm tx}}$, is first split into its real and imaginary parts to form a two-channel real-valued tensor of size $2 \times N_{\rm d} \times M_{\rm tx}$.
Then, the CNN encoder processes this tensor to extract a continuous latent representation, given by
\begin{equation}
\label{eq:Zt_def}
\mathbf{Z}_t \triangleq [\mathbf{z}_{t,0},\, \mathbf{z}_{t,1},\, \ldots,\, \mathbf{z}_{t,N-1}]^T \in \mathbb{R}^{N\times D},
\end{equation}
where $N$ is the total number of tokens per frame (whose choice will be discussed in Section~\ref{sec:experiments}) and $\mathbf{z}_{t,n}\in\mathbb{R}^{D \times 1}$ is the $n$-th token subvector ($n\in\{0,1,\dots,N-1\}$).
This step reduces the dimensionality of the CSI from $N_{\rm d} \times M_{\rm tx} \times 2$ to $N \times D$, while retaining the key angular- and delay-domain correlation features of the CSI matrix.

\textbf{CSI tokenization:}
To further compress $\mathbf{Z}_t$ into a sequence of discrete tokens, we introduce a learnable VQ codebook $\mathcal{C} = \{\mathbf{c}_0, \mathbf{c}_1, \ldots, \mathbf{c}_{K-1}\}$ containing $K$ continuous codewords. Each codeword $\mathbf{c}_k \in \mathbb{R}^{D \times 1}$ is used to quantize a subvector $\mathbf{z}_{t,n}$ in \eqref{eq:Zt_def}.
Based on this codebook, the VQ layer converts the continuous latent $\mathbf{Z}_t$ into a discrete token sequence $\mathcal{K}_t \triangleq [k_{t,0},\ldots,k_{t,N-1}] \in \{0,1,\ldots,K-1\}^N$, where $k_{t,n}\in\{0,1,\ldots,K-1\}$ is the index of the selected codeword $\mathbf{c}_{k_{t,n}}$ in the codebook $\mathcal{C}$ for the $n$-th subvector.
Note that the discretization error introduced in this procedure is explicitly accounted for in the training loss, through which the CNN encoder, decoder, and VQ codebook are jointly optimized (see details in Section \ref{sec:train_loss}).
Unlike standard VQ, which only minimizes the Euclidean distance between the subvector $\mathbf{z}_{t,n}$ and the codewords $\mathbf{c}_k$~\cite{vqvae}, ECVQ-VAE selects the index $k_{t,n}$ for each subvector $\mathbf{z}_{t,n}$ in \eqref{eq:Zt_def} by minimizing an entropy-constrained objective~\cite{shin2025ecvqvae}
\begin{equation}
  \label{eq:vq_objective}
k_{t,n} = \arg\!\!\!\!\!\!\min_{k\in\{0,\ldots,K-1\}} \left\{ -\log_2(P_k) + \frac{\lambda}{D} \|\mathbf{z}_{t,n} - \mathbf{c}_k\|_2^2 \right\}.
\end{equation}
Herein, $P_k$ is the prior probability of selecting the $k$-th codeword, which can be computed from the empirical codebook usage frequency during model training \cite{shin2025ecvqvae}.
Moreover, $\lambda$ is a Lagrange multiplier that balances the trade-off between CSI reconstruction accuracy and feedback overhead.
A smaller $\lambda$ leads to a larger discretization error (and thus lower CSI reconstruction accuracy) but a smaller feedback overhead, and vice versa.
Intuitively, the first term $-\log_2(P_k)$ is approximately the code length of the $k$-th codeword \cite{cover2006elements}, so a frequently used codeword (i.e., one with a higher prior probability $P_k$) is encoded with fewer bits.
Hence, the objective in \eqref{eq:vq_objective} tends to select high-probability codewords as long as they remain close enough to the subvector $\mathbf{z}_{t,n}$.
After tokenization, the continuous CSI representation $\mathbf{Z}_t$ is compressed into a discrete token sequence $\mathcal{K}_t$, which is used for CSI distribution prediction and entropy coding.

\vspace{-0.2em}
\subsection{Training Loss}
\label{sec:train_loss}
The ECVQ-VAE is trained in an end-to-end manner over a mini-batch $\mathcal{B}_1$ of CSI frames by minimizing a composite loss function that includes the NMSE between the ground-truth and reconstructed CSI, the discretization error brought by the VQ layer, and the entropy regularization term~\cite{shin2025ecvqvae}.
Notably, different from the standard VQ-VAE/ECVQ-VAE in \cite{vqvae,shin2025ecvqvae}, which use the mean squared error (MSE) as the reconstruction loss, we use NMSE to measure the reconstruction error, which is more suitable for CSI feedback tasks.
The overall loss function for training the ECVQ-VAE is
\begin{align}
&\mathcal{L}_{\mathrm{token}} \! =\! \frac{1}{|\mathcal{B}_1|}\!\sum_{t \in \mathcal{B}_1}\!\bigg[\mathrm{NMSE}(\mathbf{H}_t, \hat{\mathbf{H}}_t)\! -\! \frac{1+\beta}{\lambda}\mathbb{E}_{n,t}\big[\log_2 P_{k_{t,n}}\big] \nonumber \\
&\!\!\!\! + \frac{1}{ND}\sum_{n=0}^{N-1}\Big(\|\mathrm{sg}[\mathbf{z}_{t,n}]\!\! - \!\! \tilde{\mathbf{z}}_{t,n}\|_2^2
+ \beta\|\mathbf{z}_{t,n}\!\! -\! \mathrm{sg}[\tilde{\mathbf{z}}_{t,n}]\|_2^2\Big)\bigg]. \label{eq:ecvq_loss}
\end{align}
The loss function in \eqref{eq:ecvq_loss} is composed of three parts: a reconstruction error term, an entropy term, and a VQ quantization term, each weighted by its respective coefficient.
Moreover, $\mathrm{sg}[\cdot]$ denotes the stop-gradient operator \cite{vqvae}.
The term $\mathrm{NMSE}(\mathbf{H}_t, \hat{\mathbf{H}}_t) = \|\mathbf{H}_t - \hat{\mathbf{H}}_t\|_F^2 / \|\mathbf{H}_t\|_F^2$ measures the error between the ground-truth CSI $\mathbf{H}_t$ and the reconstructed CSI $\hat{\mathbf{H}}_t$.
The entropy term $-\mathbb{E}_{n}[\log_2 P_{k_{t,n}}]$ measures the average length in bits for feeding back each CSI token, and $P_{k_{t,n}}$ is the prior probability of the selected codeword, computed from the empirical codebook usage frequency during training.
The Lagrange multiplier $\lambda$ is the same hyperparameter as in \eqref{eq:vq_objective} to strike a balance between CSI reconstruction accuracy and feedback overhead.
The discretization error includes two components, the codebook loss $\|\mathrm{sg}[\mathbf{z}_{t,n}] - \tilde{\mathbf{z}}_{t,n}\|_2^2$ and the commitment loss $\|\mathbf{z}_{t,n} - \mathrm{sg}[\tilde{\mathbf{z}}_{t,n}]\|_2^2$, balanced by a hyperparameter $\beta$\footnote{According to \cite{vqvae}, the NMSE of ECVQ-VAE remains essentially unchanged for $\beta\in[0.1,2]$.
As such, following \cite{shin2025ecvqvae,vqvae}, we fix $\beta=0.25$.}.
By minimizing this loss, the encoder, decoder, and VQ codebook are jointly trained to reconstruct the CSI accurately while maintaining low feedback overhead.

\vspace{-0.2em}
\section{Stage 1: History-Aware CSI Distribution Prediction}
\label{sec:transformer}
Based on the CSI token representation in Section~\ref{sec:tokenizer},  we design a customized causal Transformer for Stage 1 to \textit{predict} the token probabilities of the next CSI frame.

\vspace{-0.2em}
\subsection{CSI Distribution Prediction with Causal Transformer}
To attain low feedback overhead, we need to predict the token probabilities of CSI frames as accurately as possible, since a more accurate CSI distribution prediction yields a shorter entropy code.
The key challenge is that the CSI tokens are correlated not only \emph{across} frames, but also \emph{within} each frame.
Therefore, both the \emph{global} position of each token in the entire sequence and its \emph{intra-frame} position are crucial for accurate token-probability prediction.
However, a conventional causal Transformer with a standard APE~\cite{vaswani2017attention} explicitly encodes the global token position only.
Although the intra-frame position is implicitly determined by the global one, the neural network has to infer this periodic structure from data, which makes it difficult to exploit the correlations among tokens at the same intra-frame position across different frames.
To address the above limitations, we design a customized causal Transformer equipped with the proposed IPE,
which explicitly encodes the intra-frame token position, thereby effectively accounting for such inter-frame correlations to achieve accurate token-probability prediction.

\subsubsection{Token Embedding and Positional Embedding}
\begin{figure}[t]
\centering
\includegraphics[width=\linewidth]{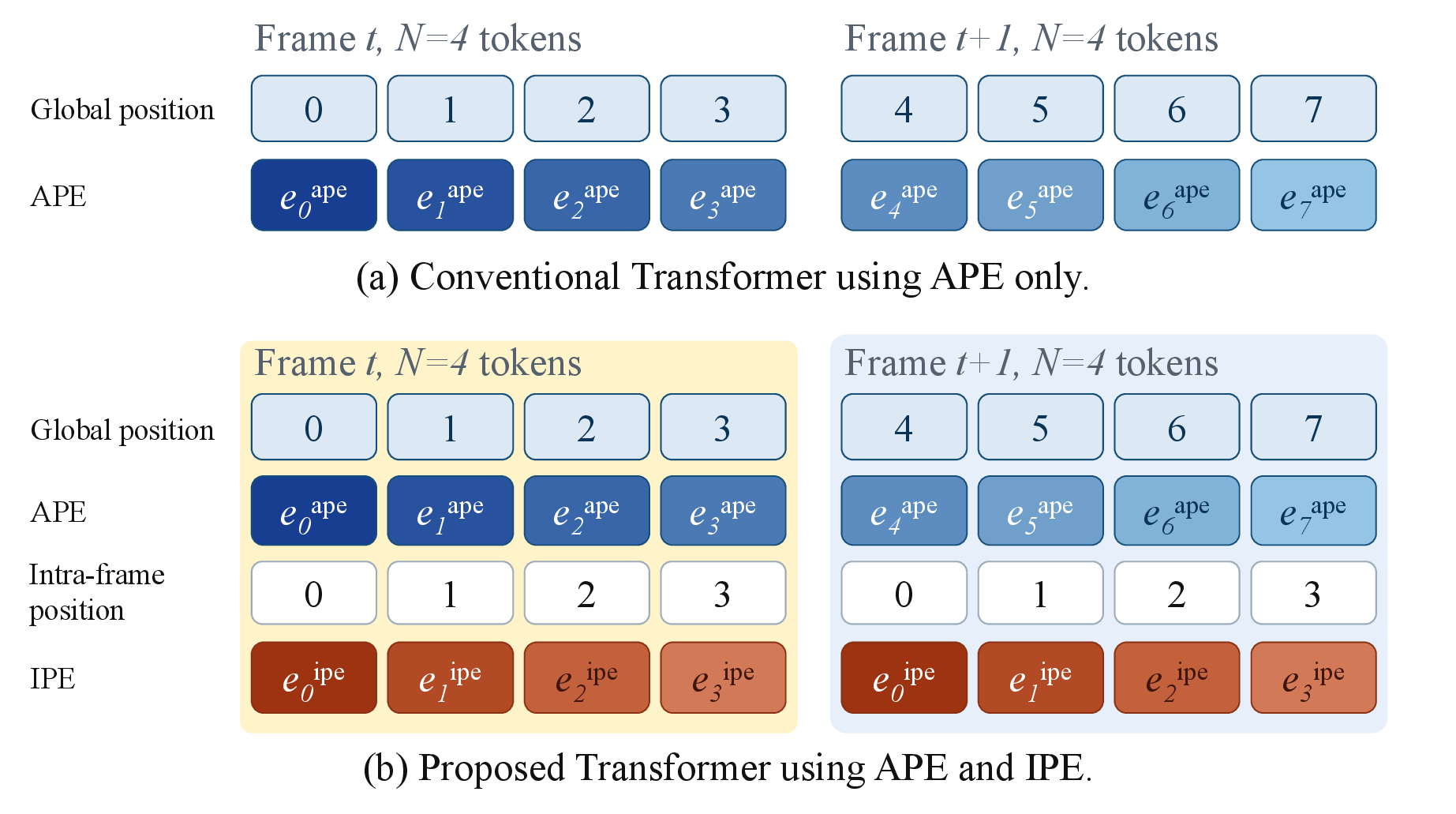}
\caption{Comparison between the APE and the proposed IPE with $N=4$ tokens per frame.
Different (or identical) colors represent different (or identical) positional embeddings.
}
\label{fig:ipe}
\vspace{-1.5em}
\end{figure}

At the current frame $t$, the available CSI token context is $\mathcal{K}_{\leq t}\triangleq [\mathcal{K}_{t-L+1},\ldots,\mathcal{K}_{t}]\in\{0,1,\ldots,K-1\}^{LN}$, where $L$ is the number of available CSI frames and $d_{\rm m}$ is the Transformer model dimension, i.e., the dimension of the token embedding and the hidden state of the Transformer.
For token {$n$} of frame {$(t+1)$}, the {$LN$}-token context is mapped to {$\mathbf{E}_{t+1,n}\in\mathbb{R}^{(LN)\times d_{\rm m}}$} via token embedding.
Notably, the Transformer cannot inherently inform the sequential order of CSI tokens \cite{vaswani2017attention}, which is crucial for modeling correlations across CSI frames and tokens.
To address this issue, we first review the standard APE and then introduce a customized IPE.
As shown in Fig.~\ref{fig:ipe},
each CSI frame is split into $N$ CSI tokens, resulting in a total of $LN$ tokens in the context window.

\textbf{Existing APE:}
In conventional Transformers, the APE is used to encode the global position of each token in the entire sequence.
Specifically, the APE is a learnable embedding table $\{\mathbf{e}^{\mathrm{ape}}_{j}\}_{j=0}^{LN-1}$, where $\mathbf{e}^{\mathrm{ape}}_{j} \in \mathbb{R}^{d_{\rm m} \times 1}$ represents the $j$-th token position in the entire sequence.

\textbf{Proposed IPE:}
To characterize this intra-frame position in addition to global position,
we design a new learnable positional embedding table $\{\mathbf{e}^{\mathrm{ipe}}_{n}\}_{n=0}^{N-1}$, where $\mathbf{e}^{\mathrm{ipe}}_{n} \in \mathbb{R}^{d_{\rm m} \times 1}$ represents the $n$-th token position within a frame and is shared across all $L$ frames.
In particular, unlike the APE, in which each token has a unique positional embedding in the entire sequence, the same IPE $\mathbf{e}^{\mathrm{ipe}}_n$ is assigned to tokens at the same intra-frame position $n$ in different CSI frames, as shown in Fig.~\ref{fig:ipe}.
The proposed IPE can capture the local intra-frame position of each token, allowing the Transformer to exploit this information for more accurate token-probability prediction.
Notably, we will show in Section~\ref{subsec:ablation} that the proposed IPE significantly improves the prediction accuracy of the token probabilities, which in turn considerably reduces the feedback overhead.

\textbf{Input embedding:}
Based on the above,
the input embedding $\mathbf{e}_j \in \mathbb{R}^{d_{\rm m} \times 1}$ of the $j$-th CSI token is given by
\begin{equation}
\mathbf{e}_j = \mathbf{e}^{\mathrm{tok}}_j + \mathbf{e}^{\mathrm{ape}}_j + \mathbf{e}^{\mathrm{ipe}}_{j \bmod N}, \quad \forall j \in \{0, 1, \ldots, LN-1\},
\end{equation}
where $\mathbf{e}^{\mathrm{tok}}_j, \mathbf{e}^{\mathrm{ape}}_j \in \mathbb{R}^{d_{\rm m} \times 1}$ denote the learnable token embedding and the standard APE for the $j$-th token, respectively.
For example, the CSI tokens at positions $j=n$ and $j=N+n$ belong to adjacent CSI frames and occupy the same intra-frame position $n$.
Hence, they share the same IPE $\mathbf{e}^{\mathrm{ipe}}_n$, but differ in their APEs (i.e., $\mathbf{e}^{\mathrm{ape}}_n$ and $\mathbf{e}^{\mathrm{ape}}_{N+n}$) since their global positions in the entire sequence are different,
and in their token embeddings (i.e., $\mathbf{e}^{\mathrm{tok}}_n$ and $\mathbf{e}^{\mathrm{tok}}_{N+n}$) since they correspond to different CSI tokens.
\subsubsection{CSI Distribution Prediction}
To predict the distribution of CSI token $n$ for the next CSI frame $\mathbf{H}_{t+1}$, the Transformer takes $\mathbf{E}_{t+1,n}$ as the input embedding through $N_{\rm layer}$ masked self-attention and feed-forward blocks, yielding hidden states $\mathbf{S}_{t+1,n}\in\mathbb{R}^{(LN)\times d_{\rm m}}$.
Next, the Transformer extracts the hidden state vector $\mathbf{s}_{t+1,n}\triangleq\mathbf{S}_{t+1,n}[LN-1,:]^T\in \mathbb{R} ^{d_{\rm m} \times 1}$.
To predict the distribution $\mathbf{p}_{t+1,n}$ of the $n$-th token in frame $(t+1)$, the hidden state $\mathbf{s}_{t+1,n}$ is then passed through a linear projection layer followed by a softmax function as follows
\begin{equation}
\label{eq:softmax}
\mathbf{p}_{t+1,n} = \mathrm{softmax}\!\left(\mathbf{W}^T \mathbf{s}_{t+1,n} + \mathbf{b}\right),\forall n \in \{0, 1, \ldots, N-1\},
\end{equation}
where $\mathbf{W} \in \mathbb{R}^{d_{\rm m} \times K}$ and $\mathbf{b} \in \mathbb{R}^{K \times 1}$ are the weight and bias of the projection layer, respectively.
Subsequently, the context window is shifted by one token to predict the next token, and this process is repeated until the distributions of all $N$ tokens of the next CSI frame are predicted.

\vspace{-0.2em}
\subsection{Training Loss}
The Transformer is trained on top of a pre-trained ECVQ-VAE.
During Transformer training, we sample a batch $\mathcal{B}_2$ of CSI-frame windows, each consisting of $L$ historical CSI frames (i.e., $LN$ CSI tokens) and the subsequent target frame.
These CSI frames are first processed by the pre-trained ECVQ-VAE to obtain the corresponding CSI token sequences.
The training loss is the average cross entropy of the true token indices as
\begin{equation}
\label{eq:transformer_loss}
\mathcal{L}_{\mathrm{transformer}} = - \frac{1}{|\mathcal{B}_2|\, N} \sum_{w \in \mathcal{B}_2} \sum_{n=0}^{N-1} \log\bigl[\mathbf{p}_{w,n}\bigr]_{k_{w,n}}.
\end{equation}
Herein, $k_{w,n}$ is the ground-truth index of the $n$-th target token in window $w$ (obtained from the pre-trained ECVQ-VAE), and $[\mathbf{p}_{w,n}]_{k_{w,n}}$ is the predicted probability assigned to that index.
By minimizing this token-level cross-entropy, the model learns the conditional distribution of each target token, thereby reducing the feedback overhead.

\vspace{-0.2em}
\section{Stage 2: Prediction-Aided CSI Entropy Coding}
\label{sec:entropy_coding}
In this section, we present the second stage for our proposed PAC, which leverages the predicted token distributions to perform entropy coding of CSI tokens at the UE and decoding at the BS.
\subsubsection{Entropy Coding of CSI Tokens at the UE}
Consider the current CSI frame at time slot $t$.
In Stage 1, based on the historical CSI tokens,
the UE predicts the conditional token probabilities $\{\mathbf{p}_{t+1,n}\}_{n=0}^{N-1}$ for the next frame $(t+1)$ using the causal Transformer proposed in Section~\ref{sec:transformer}.
Based on these predicted token distributions, the UE constructs a token-to-bitstream mapping $\mathcal{U}_{t+1}$ using an entropy coder~\cite{cover2006elements}, which converts the token sequence $\mathcal{K}_{t+1}$ into a variable-length bitstream $\mathbf{b}_{t+1} = \mathcal{U}_{t+1}(\mathcal{K}_{t+1})$.
Once the next CSI frame $\mathbf{H}_{t+1}$ arrives, in Stage~2, the UE first obtains the corresponding token sequence $\mathcal{K}_{t+1}$ using the pre-trained ECVQ-VAE.
Then, it encodes $\mathcal{K}_{t+1}$ into a variable-length bitstream $\mathbf{b}_{t+1}$ using the mapping $\mathcal{U}_{t+1}$ and feeds back $\mathbf{b}_{t+1}$ to the BS.

\subsubsection{Entropy Decoding of CSI Tokens at the BS}
The BS constructs the same entropy codebook $\mathcal{U}_{t+1}$ based on the same predicted token probabilities as in Stage 1.
Upon receiving the bitstream $\mathbf{b}_{t+1}$ sent by the UE in Stage 2,
the BS recovers $\hat{\mathcal{K}}_{t+1}=\mathcal{U}_{t+1}^{-1}(\mathbf{b}_{t+1})$,
where $\mathcal{U}_{t+1}^{-1}$ is the inverse mapping of $\mathcal{U}_{t+1}$.
Thanks to the lossless nature of entropy coding, we have $\hat{\mathcal{K}}_{t+1}=\mathcal{K}_{t+1}$.
The BS then maps $\hat{\mathcal{K}}_{t+1}$ to the quantized latent representation and uses the CNN decoder to obtain $\hat{\mathbf{H}}_{t+1}$.
However, in the conventional causal Transformer, the BS cannot construct $\mathcal{U}_{t+1}$ for all tokens at once, as elaborated next.

\begin{figure}[t]
\centering
\includegraphics[width=0.9\linewidth]{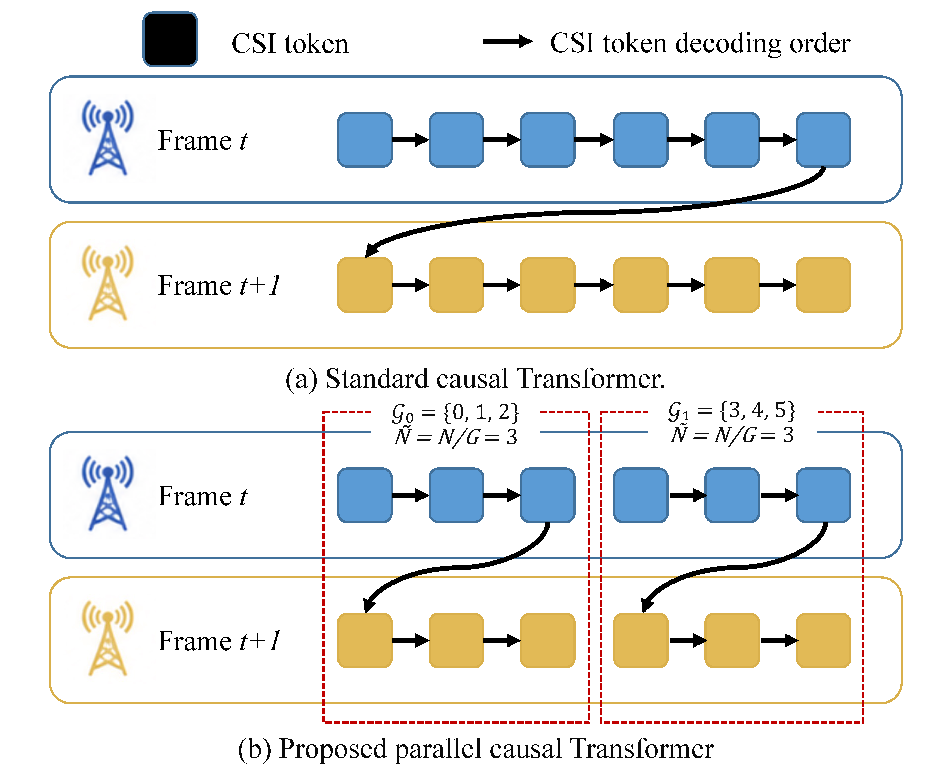}
\caption{Comparison between the standard causal Transformer and the proposed parallel causal Transformer, where $N=6$ tokens per frame are partitioned into $G=2$ groups.
}
\label{fig:decoding_flow}
\vspace{-1.5em}
\end{figure}

\textbf{Conventional sequential causal architecture:}
As shown in Fig.~\ref{fig:decoding_flow}(a),
in the conventional causal Transformer, each token in CSI frame $\mathbf{H}_{t+1}$ depends on all preceding intra-frame tokens, and therefore, the BS needs to decode the $N$ intra-frame tokens of each frame strictly in $N$ sequential steps~\cite{vaswani2017attention}.
This serial decoding procedure requires $N$ serial Transformer prediction steps per frame, resulting in high end-to-end inference latency for latency-sensitive applications.
For example, when $N=16$, the end-to-end inference latency of the conventional causal Transformer is approximately 1.4 milliseconds (ms) on the GeForce 4090 platform used in Section~\ref{sec:experiments},
which is prohibitively long for real-time communication systems.

\textbf{Proposed parallel causal architecture:}
To reduce end-to-end inference latency, we propose a new parallel causal decoding architecture.
As illustrated in Fig.~\ref{fig:decoding_flow}(b),
for each frame, we partition the $N$ tokens into $G$ consecutive groups, i.e., $\mathcal{G}_0, \ldots, \mathcal{G}_{G-1}$, each of $\tilde{N} \triangleq N/G$ tokens, with $\mathcal{G}_g = \{g\tilde{N}, g\tilde{N}+1, \ldots, (g+1)\tilde{N}-1\}, \forall g=0,1,\ldots,G-1$.
For each group $g$, we collect the $\tilde{N}$ tokens at the positions in $\mathcal{G}_g$ from each of the $L$ frames, yielding a sub-sequence of length $L\tilde{N}$ for each group.
As such, the original token sequence of length $LN$ is partitioned into $G$ sub-sequences of length $L\tilde{N}$.
For example, as shown in Fig.~\ref{fig:decoding_flow}(b), when $N=6$ and $G=2$, the first group $\mathcal{G}_0$ contains tokens at positions $\{0,1,2\}$, while the second group $\mathcal{G}_1$ contains tokens at positions $\{3,4,5\}$.
The Transformer processes the $G$ sub-sequences as a batch and decodes the next token of each group in parallel, so the $N$ sequential steps of the conventional Transformer are reduced to $\tilde{N} < N$ steps.
Note that the number of groups $G$ (equivalently, the group size $\tilde{N}$) can be adjusted to balance CSI feedback overhead and end-to-end inference latency, as discussed in \textbf{Remark~\ref{remark:gpct}}.
With this grouping scheme, the temporal correlation across frames can still be exploited for CSI compression, while the intra-frame correlation between tokens in different groups is relaxed, thereby enabling parallel decoding for fast inference.

\begin{remark}[CSI feedback overhead versus end-to-end inference latency]
  \label{remark:gpct}
  The parallel causal Transformer reduces the number of sequential decoding steps per frame from $N$ to $\tilde{N} = N/G$.
  However, the intra-frame correlation between tokens in different groups is relaxed, which may cause a slight increase in the feedback overhead due to less accurate token-probability prediction.
  For example, when $N=16$, increasing $G$ from 1 to 8 reduces the end-to-end inference latency from 1.369 ms to 0.458 ms on the GeForce 4090 platform (see Table~\ref{tab:groupk_ablation} in Section~\ref{sec:experiments}), while the feedback overhead increases from 18.30 to 19.06 bpf only, as shown in Table~\ref{tab:groupk_ablation}.
  As such, $G$ (or equivalently $\tilde{N}$) should be chosen according to the UE's mobility and the inference-latency requirement, and its effect will be numerically evaluated in Section~\ref{sec:experiments}.
\end{remark}

\vspace{-0.2em}
\section{Experimental Results}\label{sec:experiments}
\vspace{-0.2em}
\subsection{Experimental Setup}
We evaluate the proposed framework on both indoor and outdoor scenes of DeepMIMO v3.
The dataset and model configurations, together with the training parameters, are summarized in Table~\ref{tab:exp_params}.
The number of groups in the parallel causal Transformer is set to $G=4$ (with reasons specified later) unless otherwise specified.
Consecutive CSI frames are sampled along the $y$-axis in the coordinate system of the DeepMIMO dataset.
The training, validation, and testing trajectories do not spatially overlap.
All temporal windows used for training, validation, and testing are sampled within the same trajectory and do not cross trajectory boundaries, so the historical context within each window is always physically consistent with a single moving UE.
Unless otherwise specified, we assume straight UE trajectories with a constant UE speed.
For random-walk trajectories, we use a random Markov mobility model in which, at each time step, the UE moves forward with probability 0.5 or turns left or right, each with probability 0.25~\cite{3gpp_tr25942}.
All models are implemented in PyTorch and trained on a workstation equipped with two Intel Xeon Platinum 8336C central processing units (CPUs) and two GeForce 4090 graphics processing units (GPUs).

\begin{table}[t]
  \centering
  \caption{Experimental parameters.}
  \label{tab:exp_params}
  \footnotesize
  \setlength{\tabcolsep}{3pt}
  \begin{tabularx}{\columnwidth}{>{\raggedright\arraybackslash}X c c c}
  \toprule
   & \multicolumn{2}{c}{\textbf{DeepMIMO v3}} & \textbf{COST 2100} \\
  \cmidrule(lr){2-3} \cmidrule(lr){4-4}
  \textbf{Parameter} & \textbf{Indoor} & \textbf{Outdoor} & \textbf{Indoor} \\
  \midrule
  Scene & \texttt{I1\_2p5} & \texttt{O1\_3p5} & \texttt{IndoorHall\_5GHz} \\
  Carrier frequency & 2.5\,GHz & 3.5\,GHz & 5.3\,GHz \\
  Bandwidth & 50\,MHz & 50\,MHz & 20\,MHz \\
  Number of trajectories & 84 & 201 & {12000} \\
  Frames per trajectory & 1000 & 600 & {64} \\
  UE speed & 0.1\,m/s & 2\,m/s & 0.001\,m/s \\
   & \multicolumn{3}{c}{$M_{\rm tx}=N_{\rm d}=32$} \\
  & \multicolumn{3}{c}{$D=16$, $K=512$} \\
  \midrule
  Model structure (layers, width, heads) & \multicolumn{3}{c}{2, 256, 8} \\
  Context window $L$ (frames) \& dropout & \multicolumn{3}{c}{15, 0.1} \\
  Optimizer \& scheduler & \multicolumn{3}{c}{\makecell{AdamW, OneCycle}} \\
  Learning rate \& weight decay & \multicolumn{3}{c}{$3\times 10^{-4}$, $10^{-4}$} \\
  Batch size \& epochs & \multicolumn{3}{c}{160, 100} \\
  \bottomrule
  \end{tabularx}
  \vspace{-1.2em}
\end{table}

\begin{table*}[!t]
\centering
\caption{Performance comparison of different methods in both indoor and outdoor scenes, where the number (\#) of model parameters is measured in millions (M).}
\label{tab:scene_compare}
\footnotesize
\renewcommand{\arraystretch}{1.1}
\setlength{\tabcolsep}{4.9pt}
\begin{tabular}{@{}clccccc@{}}
\toprule
Scene & Scheme & Feedback overhead (bpf) $\downarrow$ & NMSE (dB) $\downarrow$ & Cosine similarity $\uparrow$ & \makecell{Inference \\latency (ms) $\downarrow$} & \# of parameters (M) $\downarrow$ \\
\midrule
\multirow{7}{*}{\textit{Indoor}}
 & History-agnostic baseline 1            & 53.8          & -16.55          & 0.989          & 0.298          & 0.81 \\
 & History-agnostic baseline 2            & 53.9          & -15.53          & 0.986          & 0.285          & 0.81 \\
 & History-agnostic baseline 3            & 72.0          & -14.97          & 0.984          & \textbf{0.276} & \textbf{0.41} \\
 & History-aware baseline 1         & 73.4          & -15.98          & 0.988          & 0.791          & 2.98 \\
 & History-aware baseline 2         & 41.6          & -15.56          & 0.986          & 0.300          & 1.07 \\
 & Proposed PAC-Seq       & \textbf{11.5} & \textbf{-17.02} & \textbf{0.990} & 0.796 & 2.81 \\
 & Proposed PAC-Paral       & 12.1          & \textbf{-17.02} & \textbf{0.990} & 0.289 & 2.81 \\
\cmidrule(l){2-7}
\multirow{7}{*}{\textit{Outdoor}}
 & History-agnostic baseline 1            & 53.1          & -13.13          & 0.975          & 0.298          & 0.81 \\
 & History-agnostic baseline 2            & 51.7          & -11.22          & 0.961          & 0.285          & 0.81 \\
 & History-agnostic baseline 3            & 72.0          &  -8.46          & 0.930          & \textbf{0.276} & \textbf{0.41} \\
 & History-aware baseline 1         & 73.4          & -12.45          & 0.972          & 0.791          & 2.98 \\
 & History-aware baseline 2         & 41.6          &  -7.29          & 0.911          & 0.300          & 1.07 \\
 & Proposed PAC-Seq       & \textbf{11.4} & \textbf{-14.53} & \textbf{0.984} & 0.788 & 2.81 \\
 & Proposed PAC-Paral       & 12.5          & \textbf{-14.53} & \textbf{0.984} & 0.293 & 2.81 \\
\bottomrule
\end{tabular}
\vspace{-1.2em}
\end{table*}

For CSI reconstruction, we consider the NMSE in dB, defined as $10 \log_{10} \left( \mathbb{E}\!\left[ \frac{\|\mathbf{H}_t - \hat{\mathbf{H}}_t\|_F^2}{\|\mathbf{H}_t\|_F^2} \right] \right)$.
We also report the cosine similarity (i.e., the correlation) between the ground-truth and reconstructed CSI, defined as
$\rho = \mathbb{E}\!\left[ \frac{\left|\mathrm{tr}\!\left(\mathbf{H}_t^H \hat{\mathbf{H}}_t\right)\right|}{\|\mathbf{H}_t\|_F \, \|\hat{\mathbf{H}}_t\|_F} \right]$.
For feedback overhead, we evaluate the average bpf over all $T$ frames and all $J$ testing trajectories.
In PAC, the entire CSI token sequence $\mathcal{K}_t$ is jointly encoded by arithmetic coding into a single frame-level bitstream $\mathbf{b}_t$ under the predicted token probabilities $\{\mathbf{p}_{t,n}\}_{n=0}^{N-1}$ from the Transformer.
The feedback overhead (in bpf), averaged over all $T$ frames and $J$ testing trajectories, is
$\frac{1}{J T}\sum_{j=1}^{J}\sum_{t=1}^{T} |\mathbf{b}^{(j)}_{t}|$, where $|\mathbf{b}^{(j)}_{t}|$ is the number of bits in the arithmetic-coded bitstream for the entire token sequence of frame $t$ on testing trajectory $j$.
For the fixed-length baselines, bpf is the total number of CSI feedback bits accumulated over the entire trajectory divided by the number of frames.
For variable-length baselines, bpf is averaged over the entire trajectory and calculated similarly to PAC.
The end-to-end inference latency is the average runtime of the entire CSI feedback process, including CSI tokenization, Transformer prediction, entropy coding, and CSI reconstruction.
For performance comparison, besides the proposed PAC with sequential decoding (\textbf{PAC-Seq}) and PAC with a customized parallel decoding (\textbf{PAC-Paral}), we consider five baselines.
\begin{itemize}
\item {History-agnostic baseline 1 (ECVQ-VAE)}~\cite{shin2025ecvqvae}: This scheme uses an ECVQ-VAE-based method for history-agnostic CSI compression, which maps each CSI frame into a sequence of CSI tokens. These CSI tokens are then arithmetic-coded using token probabilities calculated during the training process of the ECVQ-VAE, which remain fixed during testing and are agnostic to CSI history.
  \item {History-agnostic baseline 2 (VQ-VAE)}~\cite{vqvae}: This method is similar to history-agnostic baseline 1, except that it uses a VQ-VAE method for CSI compression.
  \item {History-agnostic baseline 3 (TransNet)}~\cite{cui2022transnet}: This method uses a Transformer-based model to compress each CSI frame.
  Since the original TransNet does not include quantization, a straight-through estimator-based training strategy is adopted to train the model with quantization.
  \item {History-aware baseline 1 (CsiNet with LSTM)}~\cite{wang2019csinetlstm}: This method augments a CNN-based autoencoder with an LSTM module that operates on the compressed codeword sequence to exploit temporal correlation.
\item {History-aware baseline 2 (MarkovNet)}~\cite{liu2022markovnet}: This method uses a first-order Markov model to predict the CSI frames and feeds back only the residual between the predicted and ground-truth CSI frames.
\end{itemize}

\vspace{-0.2em}
\subsection{CSI Reconstruction Accuracy versus Feedback Overhead}\label{subsec:rd_scene}
\begin{figure}[t]
\centering
\includegraphics[width=0.8\linewidth]{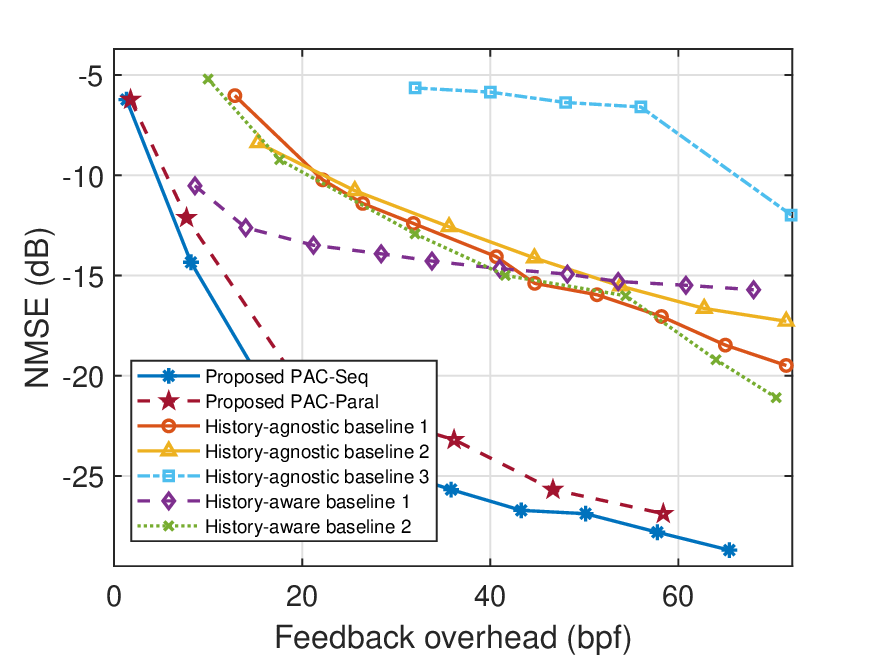}
\caption{Feedback overhead versus NMSE in the indoor scene.}
\label{fig:rd_curve_indoor}
\vspace{-1.2em}
\end{figure}

\begin{figure}[t]
\centering
\includegraphics[width=0.8\linewidth]{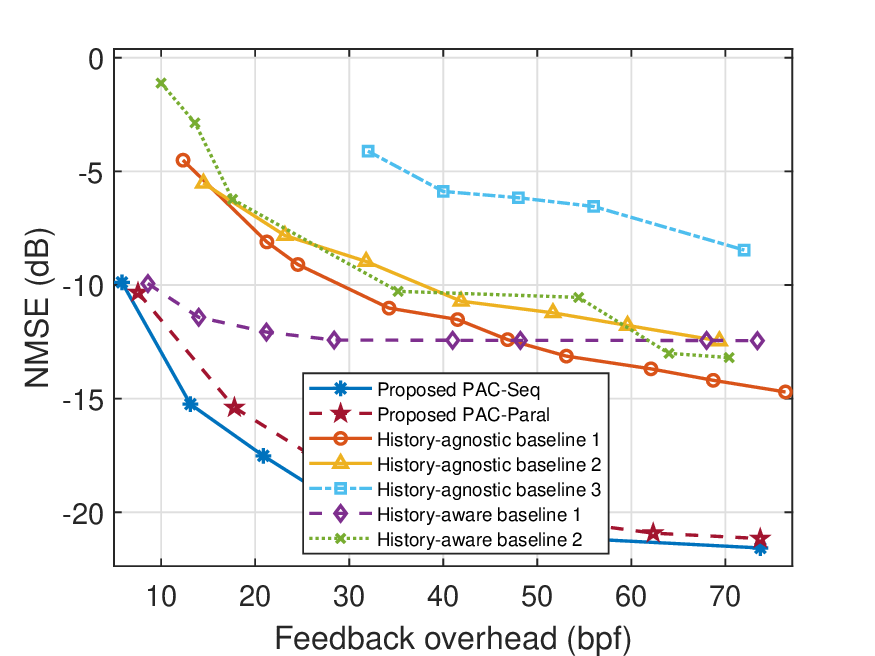}
\caption{Feedback overhead versus NMSE in the outdoor scene.}
\label{fig:rd_curve_outdoor}
\vspace{-1.2em}
\end{figure}

Figs.~\ref{fig:rd_curve_indoor} and~\ref{fig:rd_curve_outdoor} show Pareto-optimal rate-distortion frontiers\footnote{Each frontier is obtained by sweeping hyperparameters and keeping the Pareto-optimal points.
For PAC-Seq, PAC-Paral, and history-agnostic baselines 1 and 2, we sweep $\lambda\in\{10,20,40,80,200,400,800,1600\}$ and $N\in\{4,5,6,7,8,9,10,12,16,20,24,32\}$ in the indoor and outdoor scenes.
For history-agnostic baseline 3 and history-aware baselines 1 and 2, we sweep the number of quantization bits over $\{1,2,4,5,6,8\}$ and the encoder output dimension over $\{3, 7, 12, 16, 20, 25, 29, 34, 38, 43, 47, 52, 56, 60, 65\}$, respectively.} of all methods in the indoor and outdoor scenes, respectively.
It is observed that our proposed PAC dominates the rate-distortion frontier across the entire feedback-overhead range in both scenes.
The history-aware baselines generally outperform the scalar-quantization-based history-agnostic baseline (i.e., history-agnostic baseline 3) over most overhead levels.
However, for history-aware baseline 1, the NMSE saturates at sufficiently high feedback overhead, indicating that further increasing the feedback overhead provides little improvement in reconstruction gain under its LSTM-based architecture.
In contrast, history-aware baseline 2 outperforms history-aware baseline 1 at sufficiently high feedback overhead, but its CSI reconstruction accuracy degrades sharply in the low-feedback-overhead regime.
This is because the differential coding scheme used in history-aware baseline 2 relies on the accurate reconstruction of the anchor frame, which is fed back with a small number of bits in the low-feedback-overhead regime.
When the feedback overhead of the anchor frame is small, its quantization error is large and corrupts the reconstructed CSI of all subsequent frames, resulting in severely degraded NMSE.


Table~\ref{tab:scene_compare} reports the CSI feedback overhead, NMSE, cosine similarity, end-to-end inference latency, and the number of neural-network model parameters of all methods in both scenes.
Specifically, in the outdoor scene, the feedback overhead reductions of history-aware baselines are smaller than those in the indoor scene.
This is because outdoor UE speeds are much higher than indoor ones, so temporal correlation is weaker and more difficult to exploit.
Moreover, the outdoor scene exhibits a richer multipath structure, with more reflecting objects and higher-order reflection paths \cite{Alkhateeb2019DeepMIMO}.
PAC-Seq achieves the lowest feedback overhead among all schemes, requiring only $11.5$ and $11.4$ bpf in the indoor and outdoor scenes, respectively,
while achieving NMSE values of $-17.02$\,dB (indoor) and $-14.53$\,dB (outdoor).
Compared with history-agnostic baseline 1, which uses ECVQ-VAE-based CSI tokenization with history-agnostic entropy coding, PAC-Seq reduces the feedback overhead by approximately $79\%$ in both scenes while improving the NMSE by $0.47$\,dB in the indoor scene and $1.40$\,dB in the outdoor scene.
Compared with PAC-Seq, PAC-Paral substantially reduces the end-to-end inference latency from approximately $0.8$\,ms to $0.29$\,ms, while increasing the feedback overhead by only approximately $0.6$ bpf in the indoor scene and $1.1$ bpf in the outdoor scene without degrading the CSI reconstruction NMSE.

\begin{table*}[t]
\caption{Feedback overhead (bpf) and NMSE (dB) under different UE speeds.}
\label{tab:speed_rate_nmse}
\centering
\scriptsize
\renewcommand{\arraystretch}{1.15}
\setlength{\tabcolsep}{1.5pt}
\begin{tabular*}{\textwidth}{@{\extracolsep{\fill}}lcccccccccccc@{}}
\toprule
\multirow{3}{*}{Method} & \multicolumn{6}{c}{Indoor} & \multicolumn{6}{c}{Outdoor} \\
\cmidrule(lr){2-7}\cmidrule(l){8-13}
& \multicolumn{2}{c}{0.1\,m/s} & \multicolumn{2}{c}{0.2\,m/s} & \multicolumn{2}{c}{0.3\,m/s}
& \multicolumn{2}{c}{2\,m/s} & \multicolumn{2}{c}{4\,m/s} & \multicolumn{2}{c}{6\,m/s} \\
\cmidrule(lr){2-3}\cmidrule(lr){4-5}\cmidrule(lr){6-7}\cmidrule(lr){8-9}\cmidrule(lr){10-11}\cmidrule(l){12-13}
& bpf & NMSE (dB) & bpf & NMSE (dB) & bpf & NMSE (dB) & bpf & NMSE (dB) & bpf & NMSE (dB) & bpf & NMSE (dB) \\
\midrule
History-aware baseline 1 & 41.0 & \textminus12.90 & 41.0 & \textminus11.47 & 41.0 & \textminus10.50 & 41.0 & \textminus7.59 & 41.0 & \textminus6.77 & 41.0 & \textminus5.32 \\
History-aware baseline 2 & 41.6 & \textbf{\textminus14.62} & 41.6 & \textminus12.78 & 41.6 & \textminus10.52 & 41.6 & \textminus7.65 & 41.6 & \textminus7.45 & 41.6 & \textminus7.21 \\
Proposed PAC-Seq & \textbf{13.9} & \textminus14.40 & \textbf{17.9} & \textbf{\textminus14.41} & \textbf{20.8} & \textbf{\textminus14.45} & \textbf{20.4} & \textbf{\textminus16.70} & \textbf{27.6} & \textbf{\textminus16.75} & \textbf{38.8} & \textbf{\textminus16.77} \\
Proposed PAC-Paral & 14.8 & \textminus14.40 & 20.4 & \textbf{\textminus14.41} & 23.1 & \textbf{\textminus14.45} & 27.0 & \textbf{\textminus16.70} & 39.9 & \textbf{\textminus16.75} & 51.9 & \textbf{\textminus16.77} \\
\bottomrule
\end{tabular*}
\vspace{-0.8em}
\end{table*}

\begin{table*}[t]
\centering
\caption{Performance comparison under straight and random-walk (i.e., out-of-distribution) trajectories.
}
\label{tab:traj_method_compare}
\small
\begin{tabular}{ll l cc cc c}
\toprule
\multirow{2}{*}{Scene} & \multirow{2}{*}{Method} & &
\multicolumn{2}{c}{Straight} & \multicolumn{2}{c}{Random walk} &
\multirow{2}{*}{$\Delta$NMSE (dB)$\downarrow$} \\
\cmidrule(lr){4-5}\cmidrule(lr){6-7}
& & & bpf $\downarrow$ & NMSE (dB) $\downarrow$ & bpf $\downarrow$ & NMSE (dB) $\downarrow$ & \\
\midrule
\multirow{4}{*}{Indoor}
 & History-aware baseline 1  & & 73.40 & -15.18 & 73.40 & \phantom{0}-9.30  & +5.88 \\
 & History-aware baseline 2  & & 41.60 & -15.43 & 41.60 & -11.30 & +4.13 \\
 & Proposed PAC-Seq         & & 13.89          & \textbf{-16.46} & \textbf{51.75} & \textbf{-17.19} & \textbf{-0.73} \\
 & Proposed PAC-Paral        & & \textbf{13.85} & \textbf{-16.46} & 64.00          & \textbf{-17.19} & \textbf{-0.73} \\
\midrule
\multirow{4}{*}{Outdoor}
 & History-aware baseline 1  & & 80.60 & -11.31 & 80.60 & \phantom{0}-1.23 & +10.08 \\
 & History-aware baseline 2  & & 108.80 & -13.65 & 108.80 & -12.50 & +1.15 \\
 & Proposed PAC-Seq         & & \textbf{11.28} & \textbf{-13.50} & \textbf{49.40} & \textbf{-13.85} & \textbf{-0.35} \\
 & Proposed PAC-Paral        & & 12.28          & \textbf{-13.50} & 68.42          & \textbf{-13.85} & \textbf{-0.35} \\
\bottomrule
\end{tabular}
\vspace{-1.2em}
\end{table*}

\vspace{-0.2em}
\subsection{Effect of UE Speed}\label{subsec:speed}
Table~\ref{tab:speed_rate_nmse} presents the CSI feedback overhead and NMSE across three UE speed levels for both indoor and outdoor scenes.
Since history-agnostic baselines do not exploit temporal correlation, their CSI feedback overhead and NMSE are expected to be insensitive to UE speed \cite{liu2022markovnet,wang2019csinetlstm}, and these baselines are therefore omitted in the table.
As the UE speed increases, the feedback overhead of PAC grows monotonically while its CSI reconstruction NMSE remains largely unchanged in both scenes.
The reason is that a higher speed weakens the temporal correlation between consecutive CSI frames, so the CSI token probabilities predicted by the Transformer become less accurate, requiring more bits per token for the entropy coder.
However, all tokens are decoded losslessly at the BS, so the CSI reconstruction accuracy is unaffected.
History-aware baselines 1 and 2, in contrast, directly use CSI prediction to refine CSI reconstruction, and thus their NMSEs degrade sharply as the CSI prediction becomes less accurate at higher speeds.
Notably, even at the highest outdoor speed of $6$\,m/s, PAC-Seq achieves a low NMSE of $-16.77$\,dB at $38.8$\,bpf. It outperforms the best history-aware baseline by more than $9$\,dB, while requiring fewer CSI feedback bits.
As expected, the proposed PAC-Paral incurs more feedback overhead than PAC-Seq at higher speeds (see Remark~\ref{remark:gpct}).
However, its NMSE remains nearly unchanged across all speeds and is comparable to or lower than that of the best history-aware baseline.
\vspace{-0.2em}
\subsection{Out-of-distribution Trajectories}
\label{subsec:motion_type}

Table~\ref{tab:traj_method_compare} compares the CSI feedback overhead and NMSE under straight and random-walk trajectories, which serve as out-of-distribution trajectory tests because all models are trained on straight trajectories.
Note that the CSI feedback overhead and NMSE of history-agnostic baselines are insensitive to the trajectory type, so these baselines are not included.
It is observed that the CSI reconstruction accuracy of existing history-aware baselines degrades sharply under out-of-distribution trajectories and abrupt channel changes.
Such degradation is most severe for outdoor history-aware baseline 1, whose NMSE under the random-walk trajectory degrades by about $10$\,dB compared with the straight trajectory.
In contrast, since the entropy encoder sends longer bitstreams when the token-probability prediction is inaccurate, the feedback overhead increases on the random-walk trajectory while the NMSE values in all four PAC cases remain largely unchanged.

\vspace{-0.2em}
\subsection{COST 2100 Dataset}
\begin{table}[tbp]
\centering
\caption{Performance on COST 2100 (indoor).}
\label{tab:cost2100_indoor}
\begin{tabular}{lcc}
\toprule
Method & \makecell{Feedback\\overhead (bpf)} & NMSE (dB) \\
\midrule
Proposed PAC-Seq         & \textbf{14.83} & \textbf{-21.40} \\
Proposed PAC-Paral        & {16.10}        & \textbf{-21.40} \\
History-aware baseline 1 & 256            & -17.20 \\
History-aware baseline 2 & 256            & -20.00 \\
\bottomrule
\end{tabular}
\vspace{-0.5em}
\end{table}
To verify the proposed method across datasets, we also evaluate it on the COST 2100 dataset~\cite{cost2100} in Table~\ref{tab:cost2100_indoor}.
It is observed that PAC-Seq reaches an NMSE of $-21.40$\,dB at only $14.83$\,bpf.
In contrast, the history-aware baselines require $256$\,bpf but attain higher NMSE values, indicating that PAC-Seq is much more efficient in terms of CSI feedback overhead when achieving a similar CSI reconstruction accuracy.
In addition, PAC-Paral achieves the same NMSE at $16.10$\,bpf,
indicating that the proposed parallel decoding architecture can achieve low-latency CSI feedback with only a small increase in feedback overhead.
\begin{table}[tbp]
  \centering
  \caption{Ablation on the proposed IPE.
  }
  \label{tab:ipe}
  \begin{tabular}{ccccc}
    \toprule
    $N$ & Has IPE? & \makecell{Feedback\\overhead\,(bpf)} & \makecell{Parameters\\(M)} & NMSE\,(dB) \\
    \midrule
    9  & \checkmark & \textbf{11.40} & 1.880          & -14.53 \\
    9  & \texttimes & 16.01          & \textbf{1.877} & -14.53 \\
    \midrule
    16 & \checkmark & \textbf{17.80} & 1.908          & -16.59 \\
    16 & \texttimes & 32.15          & \textbf{1.904} & -16.59 \\
    \midrule
    32 & \checkmark & \textbf{36.23} & 1.974          & -19.48 \\
    32 & \texttimes & 64.19          & \textbf{1.966} & -19.48 \\
    \bottomrule
  \end{tabular}
  \vspace{-1.2em}
\end{table}
\vspace{-0.2em}
\subsection{Ablation Study}\label{subsec:ablation}

\subsubsection{IPE}
Table~\ref{tab:ipe} compares the CSI feedback overhead and the number of model parameters of the Transformer with and without the proposed IPE.
One can observe that the Transformer without IPE incurs much higher feedback overhead than the Transformer with IPE, especially when $N$ is large.
In particular, at $N=32$, IPE reduces the feedback overhead from $64.19$\,bpf to $36.23$\,bpf (a $44\%$ reduction) while increasing the number of model parameters by less than $0.5\%$.
This is because a standard APE can only capture the token position within the entire sequence, ignoring the intra-frame position of each token.
Thus, without IPE, the Transformer cannot easily exploit the intra-frame positional structure of CSI tokens,
and the predicted token probabilities become less accurate, resulting in much higher feedback overhead.
By contrast, the proposed IPE informs the position of tokens within each frame by assigning the same positional embedding to tokens at the same intra-frame position,
so the Transformer can exploit this positional structure to learn more accurate token probabilities and reduce CSI feedback overhead.
This indicates the effectiveness of IPE in improving the CSI feedback efficiency with a negligible increase in model complexity.

\subsubsection{Number of Parallel Groups \texorpdfstring{$G$}{G}}
\begin{table}[tbp]
  \centering
  \caption{Ablation on the number of parallel groups $G$ for the proposed PAC-Paral architecture.}
  \label{tab:groupk_ablation}
  \footnotesize
  \renewcommand{\arraystretch}{1.15}
  \setlength{\tabcolsep}{4pt}
  \begin{tabular}{@{}ccc@{}}
    \toprule
    $G$ & Feedback overhead (bpf) & \makecell{Inference latency (ms)} \\
    \midrule
    \multicolumn{1}{@{}c}{\textit{$N=8$}} & \multicolumn{2}{c@{}}{NMSE = -16.49\,dB} \\
    \midrule
    1 & \textbf{12.00} & 0.794 \\
    2 & 12.08          & 0.594 \\
    4 & 12.04          & 0.452 \\
    8 & 12.09          & \textbf{0.379} \\
    \midrule
    \multicolumn{1}{@{}c}{\textit{$N=16$}} & \multicolumn{2}{c@{}}{NMSE = -19.70\,dB} \\
    \midrule
    1 & \textbf{18.30} & 1.369 \\
    2 & 18.76          & 0.922 \\
    4 & 18.95          & 0.614 \\
    8 & 19.06          & \textbf{0.458} \\
    \midrule
    \multicolumn{1}{@{}c}{\textit{$N=32$}} & \multicolumn{2}{c@{}}{NMSE = -22.40\,dB} \\
    \midrule
    1 & \textbf{24.27} & 2.665 \\
    2 & 26.32          & 1.742 \\
    4 & 28.35          & 1.079 \\
    8 & 29.76          & \textbf{0.660} \\
    \bottomrule
  \end{tabular}
  \vspace{-0.8em}
\end{table}

Table~\ref{tab:groupk_ablation} summarizes the CSI feedback overhead and end-to-end inference latency of PAC-Paral for different $G$.
Because entropy coding is lossless, the BS recovers exactly the same tokens for any $G$, and thus the NMSE remains unchanged across different $G$.
As discussed in \textbf{Remark~\ref{remark:gpct}}, there is a trade-off between feedback overhead and end-to-end inference latency as $G$ changes. 
Besides the temporal correlation between inter-frame tokens, there is correlation between tokens at different intra-frame positions.
When $G$ is small, each group has more history tokens
that help capture intra-frame correlation and yield lower feedback overhead at the cost of more sequential decoding steps and higher end-to-end inference latency.
In contrast, when $G$ is large, more tokens are decoded in parallel, leading to fewer sequential decoding steps and lower end-to-end inference latency. However, the correlation between intra-frame tokens cannot be fully exploited, resulting in higher feedback overhead.
For example, when $N=16$, increasing $G$ from $1$ to $8$ reduces the end-to-end inference latency from $1.369$\,ms to $0.458$\,ms (a reduction of approximately $67\%$), while it only increases the feedback overhead by less than $1$\,bpf.
Therefore, to balance feedback overhead and end-to-end inference latency in the considered setup, we select $G=4$ as the default for PAC-Paral.

\subsubsection{Length of CSI Context Window \texorpdfstring{$L$}{L}}
\begin{figure}[t]
\centering
\includegraphics[width=0.8\linewidth]{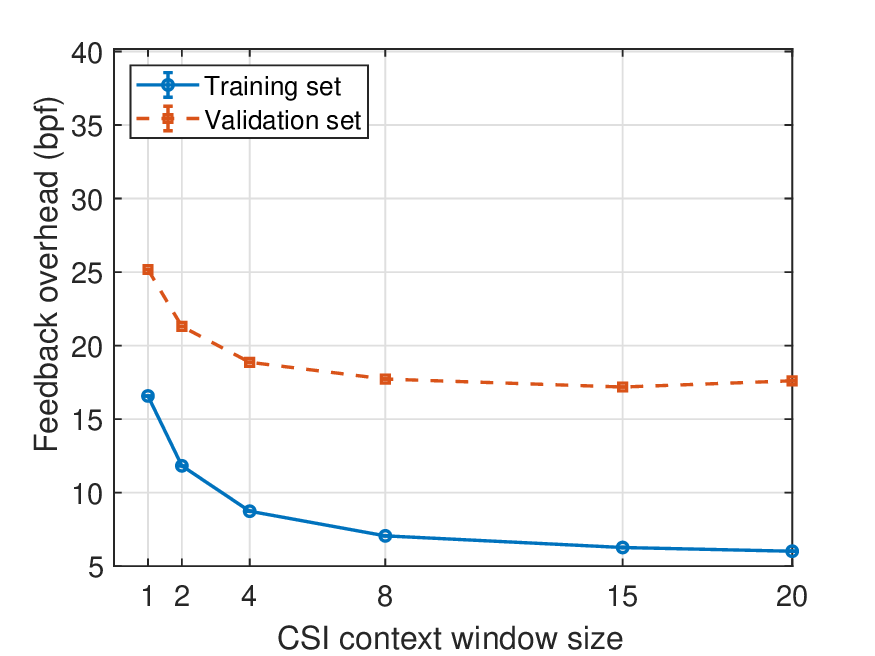}
\caption{Feedback overhead versus the context window length $L$ with $N_{\rm layer}=2$.}
\label{fig:history_bits}
\vspace{-1.2em}
\end{figure}

We plot the feedback overhead versus the context window length $L$ in Fig.~\ref{fig:history_bits}.
It is observed that the CSI feedback overhead on both the training and validation sets decreases as $L$ increases from 1 to 15.
This confirms that the model can exploit longer CSI histories for more accurate token-probability prediction and lower feedback overhead.
The validation feedback overhead reaches its minimum at about $L=15$; for $L > 15$, the gain from additional history saturates while the per-step attention cost continues to grow with $L$.
We therefore select $L=15$ as the default context window length.

\subsubsection{Sequence Model Backbone}
Table~\ref{tab:backbone_ablation} compares the NMSE, CSI feedback overhead, and end-to-end inference latency of PAC with Transformer and LSTM backbones.
One can observe that Transformer models achieve much lower end-to-end inference latency than LSTM models, especially at larger $N$.
For example, when $N=24$, the end-to-end inference latency of the Transformer is $1.85$\,ms, which is approximately $3\%$ of that of the LSTM ($60.75$\,ms) only.
Although both backbones have to decode all $N$ tokens of the next frame in a sequential manner, the Transformer can process \emph{historical} tokens in parallel, while the LSTM must process all \emph{historical} tokens sequentially.
This parallel processing of historical tokens allows the Transformer to achieve much lower end-to-end inference latency than the LSTM, especially when the historical token sequence is long (i.e., $N$ is large).
The Transformer backbone is therefore used as the default sequence model in PAC.

\begin{table}[tbp]
\centering
\caption{Performance comparison of different sequence backbones under the proposed PAC framework.}
\label{tab:backbone_ablation}
\footnotesize
\renewcommand{\arraystretch}{1.15}
\setlength{\tabcolsep}{4pt}
\begin{tabular}{@{}lccc@{}}
\toprule
Backbone & NMSE (dB) & CSI feedback (bpf) & \makecell{Inference latency (ms)} \\
\midrule
\multicolumn{4}{@{}l}{\textit{$N=4,\ \lambda=80$}} \\
\midrule
Transformer & \multirow{2}{*}{-11.25} & 8.30           & \textbf{0.60} \\
LSTM        &                           & \textbf{7.67}  & 2.15          \\
\midrule
\multicolumn{4}{@{}l}{\textit{$N=11,\ \lambda=200$}} \\
\midrule
Transformer & \multirow{2}{*}{-17.25} & \textbf{20.38} & \textbf{1.09} \\
LSTM        &                           & 25.01          & 13.72         \\
\midrule
\multicolumn{4}{@{}l}{\textit{$N=24,\ \lambda=80$}} \\
\midrule
Transformer & \multirow{2}{*}{-17.87} & \textbf{23.20} & \textbf{1.85} \\
LSTM        &                           & 26.08          & 60.75         \\
\bottomrule
\end{tabular}
\vspace{-0.8em}
\end{table}

\vspace{-0.2em}
\section{Conclusion}\label{sec:conclusion}
In this paper, we proposed PAC, a history-aware prediction-aided entropy coding framework for CSI feedback in FDD massive MIMO systems, which comprises two main stages.
In the first stage, based on the CSI history, a Transformer model predicts the distribution of next-frame CSI.
Based on such a predicted distribution, in the second stage, the UE entropy-encodes the next-frame CSI into a bitstream, which is then fed back to the BS.
Specifically, for the first stage, we proposed a novel IPE scheme to help the Transformer to exploit the intra-frame positional structure of CSI tokens.
Moreover, for the second stage, we proposed a parallel decoding architecture that allows the BS to decode multiple CSI tokens in parallel, thereby reducing the end-to-end inference latency.
Numerical results on the DeepMIMO dataset show that PAC reduces the feedback overhead by 72\% to 84\% compared to the two representative history-aware baselines, and maintains stable CSI reconstruction accuracy under random-walk trajectories where history-aware baselines degrade sharply.

\bibliographystyle{IEEEtran}

\begin{thebibliography}{10}
\providecommand{\url}[1]{#1}
\csname url@samestyle\endcsname
\providecommand{\newblock}{\relax}
\providecommand{\bibinfo}[2]{#2}
\providecommand{\BIBentrySTDinterwordspacing}{\spaceskip=0pt\relax}
\providecommand{\BIBentryALTinterwordstretchfactor}{4}
\providecommand{\BIBentryALTinterwordspacing}{\spaceskip=\fontdimen2\font plus
\BIBentryALTinterwordstretchfactor\fontdimen3\font minus
  \fontdimen4\font\relax}
\providecommand{\BIBforeignlanguage}[2]{{\expandafter\ifx\csname l@#1\endcsname\relax
\typeout{** WARNING: IEEEtran.bst: No hyphenation pattern has been}\typeout{** loaded for the language `#1'. Using the pattern for}\typeout{** the default language instead.}\else
\language=\csname l@#1\endcsname
\fi
#2}}
\providecommand{\BIBdecl}{\relax}
\BIBdecl

\bibitem{You2025NGAT}
C.~You, Y.~Cai, Y.~Liu, M.~Di~Renzo, T.~M. Duman, A.~Yener, and A.~L.
  Swindlehurst, ``Next generation advanced transceiver technologies for {6G}
  and beyond,'' \emph{IEEE J. Sel. Areas Commun.}, vol.~43, no.~3, pp.
  582--627, Mar. 2025.

\bibitem{lee2009mimo}
J.~Lee, J.-K. Han, and J.~Zhang, ``{MIMO} technologies in {3GPP} {LTE} and
  {LTE}-advanced,'' \emph{{EURASIP} J. Wireless Commun. Netw.}, vol. 2009, pp.
  302\,092:1--302\,092:10, Jul. 2009.

\bibitem{zhang2023mixedfield}
Y.~Zhang, C.~You, L.~Chen, and B.~Zheng, ``Mixed near- and far-field
  communications for extremely large-scale array: An interference
  perspective,'' \emph{IEEE Commun. Lett.}, vol.~27, no.~9, pp. 2496--2500,
  Sep. 2023.

\bibitem{zhou2026beamcoverage}
C.~Zhou, C.~You, C.~Zhou, L.~Chen, Y.~Gong, and C.~Xing, ``Low-complexity
  design for beam coverage in near-field and far-field: A {Fourier} transform
  approach,'' \emph{IEEE Trans. Wireless Commun.}, vol.~25, pp.
  21\,595--21\,610, Aug. 2026.

\bibitem{liu2026mixedfield}
T.~Liu, C.~You, C.~Zhou, M.~Wu, M.-M. Zhao, and Z.~Wang, ``Mitigating
  mixed-field interference in near-field and far-field communications: An
  antenna selection approach,'' \emph{IEEE Trans. Wireless Commun.}, vol.~25,
  pp. 20\,086--20\,101, Jul. 2026.

\bibitem{liu2023nearfieldtutorial}
Y.~Liu, Z.~Wang, J.~Xu, C.~Ouyang, X.~Mu, and R.~Schober, ``Near-field
  communications: A tutorial review,'' \emph{{IEEE} Open J. Commun. Soc.},
  vol.~4, pp. 1999--2049, Aug. 2023.

\bibitem{li2026localization}
J.~Li, C.~You, C.~Zhou, Y.~Zeng, and Z.~Feng, ``Near-field target localization:
  Effect of hardware impairments,'' \emph{IEEE Trans. Wireless Commun.},
  vol.~25, pp. 18\,885--18\,899, Jun. 2026.

\bibitem{wen2018csinet}
C.-K. Wen, W.-T. Shih, and S.~Jin, ``Deep learning for massive {MIMO} {CSI}
  feedback,'' \emph{{IEEE} Wireless Commun. Lett.}, vol.~7, no.~5, pp.
  748--751, Oct. 2018.

\bibitem{yao2026codebook}
L.~Yao, C.~You, Z.~Huang, C.~Zhou, Z.~Yang, and X.~Li, ``Codebook design for
  limited feedback in near-field {XL-MIMO} systems,'' \emph{IEEE Trans.
  Commun.}, vol.~74, pp. 6426--6442, Mar. 2026.

\bibitem{cui2022nearfield}
M.~Cui and L.~Dai, ``Channel estimation for extremely large-scale {MIMO}:
  Far-field or near-field?'' \emph{{IEEE} Trans. Commun.}, vol.~70, no.~4, pp.
  2663--2677, Apr. 2022.

\bibitem{gao2024xlcnet}
S.~Gao, P.~Dong, Z.~Pan, and X.~You, ``Lightweight deep learning based channel
  estimation for extremely large-scale massive {MIMO} systems,'' \emph{IEEE
  Trans. Veh. Technol.}, vol.~73, no.~7, pp. 10\,750--10\,754, Jul. 2024.

\bibitem{liu2022markovnet}
Z.~Liu, M.~del Rosario, and Z.~Ding, ``A {Markovian} model-driven deep learning
  framework for massive {MIMO} {CSI} feedback,'' \emph{{IEEE} Trans. Wireless
  Commun.}, vol.~21, no.~2, pp. 1214--1228, Feb. 2022.

\bibitem{wu2024dftbeamtraining}
X.~Wu, C.~You, J.~Li, and Y.~Zhang, ``Near-field beam training: Joint angle and
  range estimation with {DFT} codebook,'' \emph{IEEE Trans. Wireless Commun.},
  vol.~23, no.~9, pp. 11\,890--11\,903, Sep. 2024.

\bibitem{zhangFastBT}
Y.~Zhang, X.~Wu, and C.~You, ``Fast near-field beam training for extremely
  large-scale array,'' \emph{IEEE Wireless Commun. Lett.}, vol.~11, no.~12, pp.
  2625--2629, Dec. 2022.

\bibitem{guo2022overview}
J.~Guo, C.-K. Wen, S.~Jin, and G.~Y. Li, ``Overview of deep learning-based
  {CSI} feedback in massive {MIMO} systems,'' \emph{{IEEE} Trans. Commun.},
  vol.~70, no.~12, pp. 8017--8045, Dec. 2022.

\bibitem{guo2020csinetplus}
------, ``Convolutional neural network-based multiple-rate compressive sensing
  for massive {MIMO} {CSI} feedback: Design, simulation, and analysis,''
  \emph{{IEEE} Trans. Wireless Commun.}, vol.~19, no.~4, pp. 2827--2840, Apr.
  2020.

\bibitem{lu2020crnet}
Z.~Lu, J.~Wang, and J.~Song, ``Multi-resolution {CSI} feedback with deep
  learning in massive {MIMO} system,'' in \emph{Proc. {IEEE} Int. Conf. Commun.
  ({ICC})}, Dublin, Ireland, Jun. 2020, pp. 1--6.

\bibitem{deepcmc2021}
M.~B. Mashhadi, Q.~Yang, and D.~G\"und\"uz, ``Distributed deep convolutional
  compression for massive {MIMO} {CSI} feedback,'' \emph{{IEEE} Trans. Wireless
  Commun.}, vol.~20, no.~4, pp. 2621--2633, Apr. 2021.

\bibitem{wang2019csinetlstm}
T.~Wang, C.-K. Wen, S.~Jin, and G.~Y. Li, ``Deep learning-based {CSI} feedback
  approach for time-varying massive {MIMO} channels,'' \emph{{IEEE} Wireless
  Commun. Lett.}, vol.~8, no.~2, pp. 416--419, Apr. 2019.

\bibitem{lu2019reccsinet}
C.~Lu, W.~Xu, H.~Shen, J.~Zhu, and K.~Wang, ``{MIMO} channel information
  feedback using deep recurrent network,'' \emph{{IEEE} Commun. Lett.},
  vol.~23, no.~1, pp. 188--191, Jan. 2019.

\bibitem{liu2019dualnet}
Z.~Liu, L.~Zhang, and Z.~Ding, ``Exploiting bi-directional channel reciprocity
  in deep learning for low rate massive {MIMO} {CSI} feedback,'' \emph{{IEEE}
  Wireless Commun. Lett.}, vol.~8, no.~3, pp. 889--892, Jun. 2019.

\bibitem{cui2022transnet}
Y.~Cui, A.~Guo, and C.~Song, ``{TransNet}: Full attention network for {CSI}
  feedback in {FDD} massive {MIMO} system,'' \emph{{IEEE} Wireless Commun.
  Lett.}, vol.~11, no.~5, pp. 903--907, May 2022.

\bibitem{vqvae}
J.~Shin, Y.~Kang, and Y.-S. Jeon, ``Vector quantization for deep-learning-based
  {CSI} feedback in massive {MIMO} systems,'' \emph{{IEEE} Wireless Commun.
  Lett.}, vol.~13, no.~9, pp. 2382--2386, Sep. 2024.

\bibitem{shin2025ecvqvae}
J.~Shin, J.~Park, and Y.-S. Jeon, ``Entropy-constrained {VQ-VAE} for
  deep-learning-based {CSI} feedback,'' \emph{{IEEE} Trans. Veh. Technol.},
  vol.~74, no.~6, pp. 9870--9875, Jun. 2025.

\bibitem{li2020spatiotemporal}
X.~Li and H.~Wu, ``Spatio-temporal representation with deep neural recurrent
  network in {MIMO} {CSI} feedback,'' \emph{{IEEE} Wireless Commun. Lett.},
  vol.~9, no.~5, pp. 653--657, May 2020.

\bibitem{zhang2025futurecsi}
Y.~Zhang, A.~Alkhateeb, P.~Madadi, J.~Jeon, J.~Cho, and C.~Zhang, ``Predicting
  future {CSI} feedback for highly-mobile massive {MIMO} systems,'' in
  \emph{Proc. {IEEE} Middle East Conf. Commun. Netw. ({MECOM})}, Cairo, Egypt,
  Nov. 2025.

\bibitem{compact}
H.~Ju, S.~Jeong, S.~Kim, B.~Lee, and B.~Shim, ``Transformer-assisted parametric
  {CSI} feedback for {mmWave} massive {MIMO} systems,'' \emph{IEEE Trans.
  Wireless Commun.}, vol.~23, no.~12, pp. 18\,774--18\,787, Dec. 2024.

\bibitem{3gpp_tr36873}
{3GPP}, ``{Study on 3D channel model for LTE},'' {3rd Generation Partnership
  Project}, Technical Report TR 36.873 V12.7.0, 2017.

\bibitem{balle2017end}
J.~Ball{\'e}, V.~Laparra, and E.~P. Simoncelli, ``End-to-end optimized image
  compression,'' in \emph{Proc. Int. Conf. Learn. Represent. (ICLR)}, 2017.

\bibitem{qian2022entroformer}
Y.~Qian, X.~Sun, M.~Lin, Z.~Tan, and R.~Jin, ``Entroformer: A transformer-based
  entropy model for learned image compression,'' in \emph{Proc. Int. Conf.
  Learn. Represent. (ICLR)}, 2022.

\bibitem{Alkhateeb2019DeepMIMO}
A.~Alkhateeb, ``{DeepMIMO}: A generic deep learning dataset for millimeter wave
  and massive {MIMO} applications,'' in \emph{Proc. Inf. Theory Appl. Workshop
  (ITA)}, San Diego, CA, USA, Feb. 2019.

\bibitem{cost2100}
L.~Liu, C.~Oestges, J.~Poutanen, K.~Haneda, P.~Vainikainen, F.~Quitin,
  F.~Tufvesson, and P.~De~Doncker, ``{The COST 2100 MIMO Channel Model},''
  \emph{{IEEE} Wireless Commun.}, vol.~19, no.~6, pp. 92--99, Dec. 2012.

\bibitem{3gpp_ts38212}
{3GPP}, ``{NR; Multiplexing and channel coding},'' {3rd Generation Partnership
  Project}, Technical Specification TS 38.212 V17.4.0, 2023.

\bibitem{cover2006elements}
T.~M. Cover and J.~A. Thomas, \emph{Elements of Information Theory},
  2nd~ed.\hskip 1em plus 0.5em minus 0.4em\relax Wiley-Interscience, 2006.

\bibitem{vaswani2017attention}
A.~Vaswani, N.~Shazeer, N.~Parmar, J.~Uszkoreit, L.~Jones, A.~N. Gomez,
  {\L}.~Kaiser, and I.~Polosukhin, ``Attention is all you need,'' in
  \emph{Proc. Adv. Neural Inf. Process. Syst. ({NeurIPS})}, Long Beach, CA,
  USA, Dec. 2017, pp. 5998--6008.

\bibitem{3gpp_tr25942}
{3GPP}, ``{RF System Scenarios},'' {3rd Generation Partnership Project},
  Technical Report TR 25.942 V2.3.0, 2000.

\end{thebibliography}

\end{document}